\documentclass[twoside,preprintnumbers,amsmath,amssymb,showpacs,twocolumn]{revtex4}
\usepackage{amsfonts,amsmath,amssymb}
\usepackage[small,bf]{caption}
\usepackage{float,braket}

\newcommand{\e}{\ensuremath{\mathrm{e}}}
\newcommand{\phys}{\ensuremath{\mathrm{phys}}}

\renewcommand{\d}{\ensuremath{\mathrm{d}}}

\newcommand{\gf}{\ensuremath{\mathrm{gf}}}
\newcommand{\YM}{\ensuremath{\mathrm{YM}}}
\newcommand{\s}{\ensuremath{\mathrm{s}}}
\newcommand{\p}{\partial}
\newcommand{\h}{\ensuremath{\mathrm{h}}}
\newcommand{\GZ}{\ensuremath{\mathrm{GZ}}}

\newcommand{\sources}{\ensuremath{\mathrm{sources}}}

\begin{document}
\title{{\Large  More on the renormalization of the horizon function of the Gribov-Zwanziger action and the Kugo-Ojima Green function(s)}}

\author{D. Dudal$^a$}
    \email{david.dudal@ugent.be}
    \author{S.P. Sorella$^b$}
    \email{sorella@uerj.br}
\author{N. Vandersickel$^a$}
    \email{nele.vandersickel@ugent.be}
 \affiliation{\vskip 0.1cm
                            $^a$ Ghent University, Department of Mathematical Physics and Astronomy \\
                            Krijgslaan 281-S9, B-9000 Gent, Belgium\\\\\vskip 0.1cm
                            $^b$ Departamento de F\'{\i }sica Te\'{o}rica, Instituto de F\'{\i }sica, UERJ - Universidade do Estado do Rio de Janeiro\\
                            Rua S\~{a}o Francisco Xavier 524, 20550-013 Maracan\~{a}, Rio de Janeiro, Brasil}


\begin{abstract}
In this paper we provide strong evidence that there is no ambiguity in the choice of the horizon function underlying the Gribov-Zwanziger action. We show that there is only one correct possibility which is determined by the requirement of multiplicative renormalizability. As a consequence, this means that relations derived from other horizon functions cannot be given a consistent interpretation in terms of a local and renormalizable quantum field theory.  In addition, we also discuss that the Kugo-Ojima functions $u(p^2)$ and $w(p^2)$ can only be defined after renormalization of the underlying Green function(s).
\end{abstract}
\maketitle

\setcounter{page}{1}

\section{Introduction}
\label{intro}

In 1977, Gribov \cite{Gribov:1977wm} showed,  in a saddle point approximation,  that the restriction of the Euclidean functional integral to the Gribov region  $\Omega$ has far reaching implications for the  infrared behavior of the ghost and the gluon propagator. We recall that the region $\Omega$ is defined as the set of field configurations fulfilling the Landau gauge condition and for which the Faddeev-Popov operator,
\begin{eqnarray}
\mathcal{M}^{ab} &=&  -\p_\mu \left( \p_{\mu} \delta^{ab} + g f^{acb} A^c_{\mu} \right) \,,
\end{eqnarray}
is strictly positive. Therefore,
\begin{eqnarray}
\Omega &\equiv &\{ A^a_{\mu}, \, \p_{\mu} A^a_{\mu}=0, \, \mathcal{M}^{ab}  >0  \} \,.
\end{eqnarray}
 Later on, in a series of works,  Zwanziger \cite{Zwanziger:1989mf,Dell'Antonio:1989jn,Zwanziger:1992qr,Zwanziger:2001kw} elaborated on Gribov's  approximation,  being able to extend the previous results order by order at the quantum level.  This resulted  in an  improvement of the  Faddeev-Popov action which is now called the Gribov-Zwanziger action.  In particular, the Gribov-Zwanziger action leads to a ghost propagator which is enhanced in the infrared region,  a feature which has been confirmed by explicit  two loop calculations in \cite{Gracey:2005cx,Ford:2009ar},  which constitute a nontrivial check of the predictions of the Gribov-Zwanziger formalism.\\
 \\
Recently, it has been claimed \cite{Kondo:2009wk,Kondo:2009gc,Kondo:2009ug} that the Gribov-Zwan\-ziger action is plagued by a certain ambiguity. Depending on the choice of the  so called horizon function  \cite{Zwanziger:1989mf}, different results for the ghost propagator might be found,  namely: an enhanced or a non-enhanced ghost, according to \cite{Kondo:2009wk,Kondo:2009gc,Kondo:2009ug}. Also, the recent lattice results seem to point towards a non-enhanced ghost\footnote{However, let us mention that no unanimous consensus on this matter has yet been reached, see e.g.~\cite{Maas:2009se,Maas:2009ph,Fischer:2008uz,Oliveira:2007tr,Silva:2006av} for possible other opinions on the infrared behavior of the ghost propagator.} \cite{Cucchieri:2007md,Cucchieri:2008fc,Bornyakov:2008yx,Bogolubsky:2009dc,Maas:2008ri}. A natural conclusion would seem to be that one should take the horizon function which leads to the non-enhanced ghost. However, this does not agree with the original results by Gribov and Zwanziger and therefore asks for an explanation.\\
\\
In this paper, we would like to clear the situation. We shall show that there is no ambiguity in the choice of the horizon function. The correct form of the horizon function is the one originally constructed by Zwanziger \cite{Zwanziger:1992qr}, and is clearly dictated by the renormalization properties  of the Gribov-Zwanziger action. We stress that renormalization is of paramount importance for defining meaningful Green functions. A dynamical improvement of this Gribov-Zwanziger action, consistent with the renormalization, consequently allows to obtain the non-enhanced ghost, as discussed in previous work \cite{Dudal:2007cw,Dudal:2008sp}, giving results compatible with other analytical approaches \cite{Fischer:2008uz,Boucaud:2008ji,Boucaud:2008ky,Aguilar:2008xm,Aguilar:2009nf}, based on the Schwinger-Dyson formalism, see also \cite{Huber:2009tx}.\\
\\
As a corollary of the present analysis, we shall elaborate on the meaning of the Kugo-Ojima functions $u(p^2)$ and $w(p^2)$ defined as follows \cite{Kugo:1979gm,Kugo:1995km}
\begin{multline} \label{ww}
\int \d^d x \d^d y e^{i p(x-y) } \Braket{ (g f_{abc} A_\mu^b c^c) (x)  (g f_{ak\ell} A_\nu^k c^\ell ) (y) }_{1PI}  \\
= g_{\mu\nu} u(p^2) + \frac{p_\mu p_\nu}{p^2} w(p^2)
\end{multline}
and their meaning at the level of renormalization, shedding more light on certain claims in \cite{Kondo:2009wk,Kondo:2009gc,Kondo:2009ug}. We shall also discuss about several results obtained in the literature \cite{Kondo:2009wk,Kondo:2009gc,Kondo:2009ug}, where the other choice of the horizon function was investigated. As we shall show that this particular choice cannot be consistently introduced within a local and renormalizable framework, it follows that  its connection with the ghost propagator's behaviour, encoded in the quantities $u(p^2)$  and $w(p^2)$, cannot be properly defined at the quantum level. \\
\\
The paper is organized as follows. In section II, after giving a detailed account of the aforementioned claims \cite{Kondo:2009wk,Kondo:2009gc,Kondo:2009ug} about the existence of a possible ambiguity in the choice of the horizon term, we shall show that there is only one possible horizon function which is dictated by the renormalization properties of the Gribov-Zwanziger action. This shall be extensively discussed in sections III and IV, thereby showing that there is in fact no such ambiguity. In section V, we shall spend some words on the Kugo-Ojima Green function \eqref{ww} and on an alternative one which, in contrast with expression \eqref{ww}, turns out to be  renormalizable and closely related to the horizon function. We shall end this paper with a discussion, by emphasizing our possible explanation for the recent lattice results. In the Appendix A, we have provided the complete proof of the renormalization of the Gribov-Zwanziger action, needed for the analysis performed in sections III-V, whereby we have also taken the opportunity to correct a small mistake concerning a minor statement in our previous works \cite{Dudal:2008sp,Dudal:2005na,Dudal:2009zh}. However, none of our results \cite{Dudal:2007cw,Dudal:2008sp,Dudal:2009zh} are affected by this.

\section{Survey of the issue concerning a possible ambiguity in the choice of the horizon function}
 In \cite{Zwanziger:1992qr}, it has been shown that the restriction to the Gribov region $\Omega$ can be established by adding the following nonlocal term
\begin{eqnarray}
S_\h&=&\int \d^d x\, h(x) \,,
\end{eqnarray}
to the standard Yang-Mills action.
\begin{eqnarray}
 S_\YM + S_\gf\,,
\end{eqnarray}
with $S_{\YM}$ the classical Yang-Mills action and $S_\gf$ the Landau gauge fixing
\begin{eqnarray}
S_{\YM} &=& \frac{1}{4}\int \d^d x F^a_{\mu\nu} F^a_{\mu\nu}\;, \nonumber\\
S_{\gf} &=& \int \d^d x\,\left( b^a \p_\mu A_\mu^a +\overline c^a \p_\mu D_\mu^{ab} c^b \right)\,.
\end{eqnarray}
The nonlocal expression $h(x)$ is called the horizon function. In \cite{Kondo:2009wk}, it has been argued that there are in fact two possible choices for the horizon functions, namely
\begin{equation}\label{h1}
h_1(x) =  \gamma^4  \int \d^d y  g^2 f^{akc} A^k_\mu(x) (\mathcal M^{-1})^{ab} (x,y) f^{b\ell c} A^\ell_\mu (y)\,,
\end{equation}
or
\begin{eqnarray}
h_2(x) &=& \gamma^4 \int \d^d y \; D_\mu^{ac}(x) (\mathcal M^{-1})^{ab}(x,y) D_\mu^{bc} (y) \,,
\end{eqnarray}
with $\mathcal M$ the Faddeev-Popov operator
\begin{eqnarray}
\mathcal M^{ad} &=& -\p_\mu D_\mu^{ad} \;,
\end{eqnarray}
and
\begin{eqnarray}
D_\mu^{ab} &=& \p_\mu \delta^{ab} +  g f^{akb} A^k_\mu
\end{eqnarray}
the covariant derivative. The parameter $\gamma$ is not free, but is fixed by a gap equation, known as the horizon condition, which reads
\begin{eqnarray}\label{horizoncondition}
\braket{h(x)} &=& d (N^2 -1) \,,
\end{eqnarray}
with $d$ the number of space-time dimensions. This condition ensures that the domain of integration in the Feynman path integral has been restricted to field configurations belonging to the Gribov horizon $\Omega$. In this paper, we shall motivate that there is only one possible horizon function, the one which was originally intended by Zwanziger\footnote{See also \cite{Zwanziger:1993dh}.} \cite{Zwanziger:1992qr}, given by
\begin{multline}\label{danhorizon}
S_\h = \lim_{\theta \to 0}  \int \d^d x h_3 (x) = \lim_{\theta \to 0}  \int \d^d x \int \d^d y  \; \\
\times \left( D_\mu^{ac}(x) \gamma^2(x) \right)(\mathcal M^{-1})^{ab}(x,y) \left( D_\mu^{bc} (y) \gamma^2(y) \right) \;,
\end{multline}
whereby $\gamma(z)$ is defined through
\begin{eqnarray}\label{thetadep}
\gamma^2 (z) &=& \e^{i \theta z} \gamma^2 \;.
\end{eqnarray}
The $\lim_{\theta \to 0}$ operation corresponds to replacing the space time dependent $\gamma^2(z)$ with the constant Gribov parameter $\gamma^2$. We observe that this horizon function shares a great resemblance with $h_1(x)$ and $h_2(x)$. It is worth to point out here that the limit, $\lim_{\theta \to 0}$, in expression \eqref{danhorizon} is meant to be taken after an appropriate localization of the horizon function, a point which we shall outline in detail in what follows.\\
\\
Let us try to explain here how it is possible that some doubts have arisen about which horizon function is the correct one. Firstly, one can easily see that setting $\gamma(x)$ in $h_3(x)$ immediately equal to a constant parameter, $\gamma(x)\equiv\gamma$, which agrees with switching the limit $\lim_{\theta \to 0}$ and the integration signs, we obtain the horizon function $h_1(x)$. One can therefore appreciate that the difference between $h_1(x)$ and $h_3(x)$ is very subtle. Comparing $h_2(x)$ with $h_3(x)$, it is apparent that the space time dependent function $\gamma^2(y)$ has been pulled out of the horizon function, an operation which is not completely justified, due to the nonlocal character of the kernel  $(\mathcal M^{-1})^{ab}(x,y)$. A second source of a possible ambiguity can also arise when we try to localize the horizon function $h_3(x)$. Indeed, we can localize the action $ S_\YM + S_\gf + S_\h$ by introducing some extra fields. Looking at the following standard formula for Gaussian integration for bosonic fields
\begin{multline}\label{alg2}
 C \det A^{-1} \exp  \int \d^d x \d^d y \   J^a_\varphi(x) (A^{-1})^{ab}(x,y) J^b_{\overline{\varphi}}(y) \\
 = \int [\d \varphi][\d \overline{\varphi}] \exp\Bigl[ \int \d^d x \d^d y - \overline{\varphi}^a(x) A^{ab} (x,y)\varphi^b(y) \\
 + \int \d^4 x \ (\varphi^a J^a_\varphi(x)  + \overline{\varphi}^a(x) J^a_{\overline{\varphi}} (x)) \Bigr] \;,
\end{multline}
we observe that we can get rid of the inverse of the Faddeev-Popov operator in $h_3(x)$ by introducing new fields. For every index $i$, defined by $\ldots_i^a = \ldots^{ac}_\mu$, we can write for $h_3(x)$
\begin{align}
&\exp\left( - \int \d^d x h_3(x) \right) = \prod_{i = 1}^{d (N^2 +1)} \det (-\mathcal M)\int [\d \varphi][\d \overline{\varphi}] \nonumber\\
&  \exp \Biggl( \lim_{\theta\to 0} \Bigl[   \int \d^d x \int \d^d y \overline \varphi_i^a (x) \mathcal M^{ab}(x,y)   \varphi^b_i (y) \nonumber\\
& + \int \d^d x \left( D_i^{a} (x) \gamma^2(x) \right) \varphi_i^a(x) +  \left( D_i^a (x) \gamma^2(x) \right) \overline \varphi_i^a(x) \Bigr] \Biggr) \;,
\end{align}
whereby we have introduced a pair of complex conjugate bosonic fields $\left(\overline \varphi_\mu^{ac}, \varphi_\mu^{ac}\right)$ $=$ $\left(\overline \varphi_i^{a}, \varphi_i^{a}\right)$. We can then also lift the determinants $ \det (-\mathcal M)$ into the exponential by introducing a pair of Grassmann fields $\left( \overline \omega_\mu^{ac},\omega_\mu^{ac} \right)$ $=$ $\left( \overline \omega_i^{a},\omega_i^{a} \right)$. Making use of the standard Gaussian formula for Grassmann variables
\begin{multline}\label{alg1}
	C \left(\det A\right) \exp\left(  -\int \d^d x \d^d y \   J_\omega^a(x) (A^{-1})^{ab}(x,y) J_{\overline{\omega}}^b(y) \right) \\
= \int [\d\omega][\d\overline{\omega}] \exp\left[ \int \d^d x \d^d y \ \overline{\omega}^a(x) A^{ab} \omega^b(y) \right. \\
\left.+ \int \d^4 x \ (J^a_\omega(x) \omega^a(x) + \overline{\omega}^a(x) J_{\overline{\omega}}^a (x)) \right] \;,
\end{multline}
whereby we set the sources $J_\omega^a $ and $J_{\overline{\omega}}^b$ equal to zero, we obtain
\begin{multline*}
\exp\left( - \int \d^d x h_3(x) \right) = \prod_{i = 1}^{d (N^2 +1)}  \int [\d\omega][\d\overline{\omega}][\d \varphi][\d \overline{\varphi}]  \\
\exp \Bigl[  \int \d^d x \int \d^d y  \left( \overline \varphi_i^a (x) \mathcal M^{ab}(x,y)   \varphi^b_i (y) \right.\\
 \hspace{4cm}\left.- \overline \omega_i^a (x) \mathcal M^{ab}(x,y)   \omega^b_i (y)\right)\\
  +  \lim_{\theta\to 0} \int \d^d x \left( D_i^{a} (x) \gamma^2(x) \right) \varphi_i^a(x) \\
 +  \left( D_i^a (x) \gamma^2(x) \right) \overline \varphi_i^a(x) \Bigr] \;.
\end{multline*}
The new localized action thus becomes
\begin{eqnarray}\label{SGZphys}
S_\GZ &=& S_0' +  S_\gamma' \;,
\end{eqnarray}
with
\begin{equation}\label{SGZ}
S_0' = S_\YM + S_\gf   + \int \d^d x\left( \overline \varphi_\mu^{ac} \p_\nu D_\nu^{ab} \varphi_\mu^{bc}  - \overline \omega_\mu^{ac} \p_\nu D_\nu^{ab} \omega_\mu^{bc}   \right) \,,
\end{equation}
and with
\begin{eqnarray}\label{hor1a}
S_\gamma' &=& -  \lim_{\theta \to 0} \int \d^d x\left[ \left( D_\mu^{ac} (x) \gamma^2(x) \right) \varphi_\mu^{ac}(x) \right. \nonumber\\
&& \hspace{3cm} \left. + \left( D_\mu^{ac} (x) \gamma^2(x) \right) \overline \varphi_\mu^{ac}(x)  \right]\nonumber   \\
&=&  \lim_{\theta\to 0} \int \d^d x \;  \gamma^2(x) D_\mu^{ca} (\varphi_\mu^{ac}(x) + \overline \varphi_\mu^{ac}(x))  \nonumber\\
&=&  \gamma^2 \int \d^d x  \; D_\mu^{ca} (\varphi_\mu^{ac}(x) + \overline \varphi_\mu^{ac}(x))\;. \label{bb1}
\end{eqnarray}
Notice that, as already remarked, the limit $\theta\to 0 $,  in equation \eqref{bb1} has been performed after localization. As one can see from \eqref{thetadep}, taking this limit is equivalent with setting $\gamma^2(x)$ equal to the constant $\gamma^2$. As at the level of the action, total derivatives are always neglected, $S_\gamma'$ becomes
\begin{eqnarray}\label{h1bis}
S_\gamma '&=&   \gamma ^{2}  \int \d^d x g  f^{abc}A_\mu^a \left( \varphi_\mu^{bc} +  \overline \varphi_\mu^{bc} \right) \,.
\end{eqnarray}
From this expression we can easily understand why certain doubts have arisen. Starting from the first horizon function $h_1(x)$ given in \eqref{h1} and undertaking the same procedure, we would end up with exactly the same action $S_\gamma'$. This can be understood as we have neglected the total derivatives. In conclusion, although the local actions derived from $h_1(x)$ and $h_3(x)$ are the same, at the nonlocal level they are clearly different, which also follows from \cite{Kondo:2009wk}. This is important when one is doing manipulations at the level of the nonlocal action as has been done in \cite{Kondo:2009wk}. \\
\\
Let us now translate the nonlocal horizon condition \eqref{horizoncondition} into a local version \cite{Zwanziger:1992qr}. The local action $S_\GZ$ and the nonlocal action $S_\YM + S_\gf + S_\h$ are related as follows,
\begin{multline}
    \int [\d A] [\d b] [\d c][ \d \overline{c}] \e^{- (S_\YM + S_\gf + S_\h)} \nonumber\\
    = \int [\d A ][\d b][ \d c ][\d\overline{c}][ \d \varphi][ \d\overline{\varphi}][ \d \omega][ \d \overline{\omega}] e^{-S_\GZ} \;.
\end{multline}
Next, we take the partial derivative of both sides with respect to $\gamma^2$ so we obtain,
\begin{eqnarray}
-2 \gamma^2 \braket{h} &=& \braket{g f^{abc} A^a_{\mu} ( \varphi^{bc}_{\mu} +  \overline{\varphi}^{bc}_{\mu})}  \;,
\end{eqnarray}
for both horizon functions $h_1(x)$ and $h_3(x)$. We recall that $\braket{\p_\mu \varphi^{aa}} = 0$ and $\braket{\p_\mu \overline \varphi^{aa}} = 0$, meaning that both horizon functions $h_1$ and $h_3$ give rise to the same local horizon condition.  Using these expressions and assuming that $\gamma \not= 0$, we can rewrite the horizon condition \eqref{horizoncondition}
\begin{eqnarray}
\braket{g f^{abc} A^a_{\mu} ( \varphi^{bc}_{\mu} +  \overline{\varphi}^{bc}_{\mu})}   + 2 \gamma^2 d (N^2 -1)   = 0  \; . \label{horizoncondition2b}
\end{eqnarray}
By adding the vacuum term
\begin{eqnarray}\label{vac}
\int \d^d x \;  \gamma^4 d (N^2 - 1 )
\end{eqnarray}
to $S_\gamma'$, we can write the horizon condition as
\begin{eqnarray}\label{gapgamma}
\frac{\p \Gamma}{\p \gamma^2} &=& 0\;,
\end{eqnarray}
with $\Gamma$ the quantum action defined as
\begin{eqnarray}
\e^{-\Gamma} &=& \int [\d\Phi] \e^{-S_\GZ} \;,
\end{eqnarray}
where $\int [\d\Phi]$ stands for the integration over all the fields.\\
\\
For the Gribov-Zwanziger action to be renormalizable, it necessary to perform a shift over the field $\omega^a_i$, see \cite{Zwanziger:1992qr},
\begin{multline}\label{shift}
\omega^a_i (x) \to  \omega^a_i (x) \\
+ \int \d^d z (\mathcal M^{-1})^{ad} (x,z) g f_{dk\ell} \p_\mu [ D_\mu^{ke} c^e (z) \varphi^\ell (z)]\;,
\end{multline}
so that the action becomes
\begin{eqnarray}\label{SGZecht}
S_\GZ &=& S_0 +  S_{\gamma} \,,
\end{eqnarray}
whereby $S_0'$ has been replaced by $S_0$
\begin{eqnarray}\label{S0}
S_0 &=& S_0' + \int \d^d x \left( - g f^{abc} \p_\mu \overline \omega_i^a    D_\mu^{bd} c^d  \varphi_i^c \right)\,,
\end{eqnarray}
and the vacuum term is now included in $S_\gamma$
\begin{eqnarray}
S_{\gamma}&=& S_\gamma' + \int \d^d x  \; \gamma^4 d (N^2 - 1 ) \;.
\end{eqnarray}
The renormalizability of $S_\GZ$ has been proven in the Appendix A. We would like to stress that this is far from being trivial, especially since no new parameter is needed to take into account the divergences of the vacuum term we introduced by hand in equation \eqref{vac}. In addition, the algebraic formalism employed in the Appendix A also gives a more clean argument why the extra term appearing in equation \eqref{S0} is necessary, without the need of performing the nonlocal shift \eqref{shift}.

\section{The composite operators, part I}
We shall provide a strong argument which illustrates that only the horizon function $h_3(x)$ possesses a clear meaning at the quantum level. For this, we first need to demonstrate the following equality
\begin{eqnarray} \label{head}
\Braket{ f(A) (\mathcal M^{-1})^{ab}(x,y)  } &=& \Braket{ f(A) c^a (x) \overline c^b(y)  } \nonumber\\
&=& - \frac{1}{N_c}\Braket{ f(A) \omega^a_i (x) \overline \omega^b_i(y)  }\;,
\end{eqnarray}
where $f(A)$ stands for an arbitrary quantity depending on the gauge fields $A^a_\mu$, and $N_c$ is the number of colors.
Let us start with
\begin{eqnarray}
 \Braket{ f(A) c^a (x) \overline c^b(y)  } &=& \int [\d \Phi] f(A) c^a (x) \overline c^b(y) e^{-S_\GZ}\;.
\end{eqnarray}
We can rewrite this expression by adding corresponding sources for the ghost and the gluon fields,
\begin{multline} \label{eq}
 \Braket{ f(A) c^a (x) \overline c^b(y)  } \\
 = \left.  f\left(\frac{\delta}{\delta J_A} \right)\frac{ \delta}{\delta J_{\overline c}^b(y)} \frac{ \delta}{\delta J_c^a(x)}  \int [\d \Phi]   e^{-S_\GZ + S_\sources} \right|_{\sources = 0}\;,
\end{multline}
with
\begin{multline}
S_\sources \\= \int \d^d x \left\{ \begin{bmatrix}
J_{\omega_i}^a & J_c^a
\end{bmatrix}
\begin{bmatrix}
\omega^a_i \\ c^a
\end{bmatrix}  + \begin{bmatrix}
\overline \omega^a_i & \overline c^a
\end{bmatrix}
\begin{bmatrix}
\overline J_{\overline \omega_i}^a \\ \overline J_{\overline c}^a
\end{bmatrix}   + J_A A\right\}\;,
\end{multline}
and with the full expression for $S_\GZ$ given in \eqref{SGZecht}. We can perform the integration over the ghosts $c$, $\overline c$, $\omega$, $\overline \omega$ as this is just a Gaussian integration. The relevant piece of the action is given by
\begin{equation}
S_\GZ = \int \d^d x \begin{bmatrix}
\overline \omega_i^a & \overline c^a
\end{bmatrix}\underbrace{
\begin{bmatrix}
 -\mathcal M^{ab} & g f_{ak\ell} \p_\mu ( \varphi_i^\ell D_\mu^{kb} ) \\
0 &  \mathcal M^{ab}
\end{bmatrix}}_{K^{ab}}
\begin{bmatrix}
\omega_i^c  \\ c^d
\end{bmatrix}  + \ldots \;.
\end{equation}
Making use of the formula \eqref{alg1} gives
\begin{multline}
 \Braket{ f(A) c^a (x) \overline c^b(y)  }   f\left(\frac{\delta}{\delta J_A} \right)\frac{ \delta}{\delta J_{\overline c}^b(y)} \frac{ \delta}{\delta J_c^a(x)}  \\
 = \left.\int [\d \Phi]   \exp \left\{  \begin{bmatrix}
J_{\omega_i}^a & J_c^a
\end{bmatrix}  (K^{-1})^{ab}   \begin{bmatrix}
\overline J_{\overline \omega_i}^b \\ \overline J_{\overline c}^b
\end{bmatrix}  + \ldots \right\}  \right|_{\sources = 0} \;.
\end{multline}
The matrix $(K^{-1})^{ab}$ can be computed as
\begin{eqnarray}
(K^{-1})^{ab} &=&
 \begin{bmatrix}
-(\mathcal M^{-1} )^{ab} & \chi\\
0 & (\mathcal M^{-1} )^{ab}
\end{bmatrix}\;,
\end{eqnarray}
whereby $\chi$ is some function of the relevant fields. From \eqref{eq}, we now see that
\begin{eqnarray}
 \Braket{ f(A) c^a (x) \overline c^b(y)  } &=&    \int [\d \Phi] f(A) (\mathcal M^{-1} )^{ab} e^{-S_\GZ }\;,
\end{eqnarray}
while
\begin{multline}
 \Braket{ f(A) c^a (x) \overline c^b(y)  } = - \frac{1}{N_c} f\left(\frac{\delta}{\delta J_A} \right)\frac{ \delta}{\delta J_{\overline \omega_i}^b(y) } \frac{ \delta}{\delta J_{\omega_i}^a(x)}\nonumber\\
\\ \left.    \int [\d \Phi]    \exp \left\{  \begin{bmatrix} J_{\omega_i}^a & J_c^a \end{bmatrix}  (K^{-1})^{ab}   \begin{bmatrix}
\overline J_{\overline \omega_i}^b \\ \overline J_{\overline c}^b
\end{bmatrix}  + \ldots \right\}  \right|_{\sources = 0} \nonumber\\
\\ = -\frac{1}{N_c}\Braket{ f(A) \omega_i^a (x) \overline \omega_i^b(y)  }\;,
\end{multline}
which proves relation \eqref{head}.

\subsection{The horizon function $h_3$}
We shall now prove that the expectation value $\Braket{h_3(x)}$, see equation \eqref{danhorizon}, is renormalizable. We can apply the formula \eqref{head} to the horizon condition $\Braket{h_3(x) }$, yielding
\begin{align}\label{ident}
&\int \d^d x \Braket{h_3(x)} = \lim_{\theta \to 0} \int \d^d x \int \d^d y  \nonumber\\
&\hspace{1.5cm}\Braket{ \left( D_\mu^{ac}(x) \gamma^2(x) \right)  (\mathcal M^{-1})^{ab}(x,y) \left( D_\mu^{bc} (y) \gamma^2(y) \right) }\nonumber  \\
&\hphantom{\int \d^d x \Braket{h_3(x)}} =  -\lim_{\theta\to 0} \int \d^d x \int \d^d y  \nonumber\\
&\hspace{2cm} \Braket{ \left( D_\mu^{ac}(x) \gamma^2(x) \right) \omega_i^a (x) \overline \omega_i^b (y)  \left( D_\mu^{bc} (y) \gamma^2(y)\right) } \nonumber  \\
&\hphantom{xxxxx}= -\frac{1}{N_c}\gamma^4 \int \d^d x \int \d^d y \braket{ ( D_\mu  \omega_i)^a (x) ( D_\mu \overline \omega_i)^a (y) } \;.
\end{align}
In order to check whether this horizon term $h_3(x)$ is well defined, the correlator $\braket{ ( D_\mu  \omega_i)^a (x) ( D_\mu \overline \omega_i)^a (y) }$ including the composite operators $( D_\mu  \omega_i)^a(x)$ and $( D_\mu \overline \omega_i)^a(y)$ should be renormalizable. In fact, this turns out to be the case, as shown in the Appendix A, where a detailed account of the algebraic renormalization of the Gribov-Zwanziger action has been provided. Moreover, we have also proven that the action $\Sigma_\GZ$ (see equation \eqref{brstinvariant}) is renormalizable. From this, one can immediately obtain the correlator $\braket{ ( D_\mu  \omega_i)^a (x) ( D_\mu \overline \omega_i)^a (y) }$ by deriving the action $\Sigma_\GZ$ \eqref{brstinvariant} with respect to the sources\footnote{We recall here that in order to consistently discuss composite operators at the quantum level, they need to be introduced into the theory by means of suitable sources.} $N^{a}_{i}(x)$ and $U^{a}_{i}(y)$,
\begin{eqnarray}\label{2terms}
&&\int [\d \Phi] \left. \frac{\delta }{\delta N^{a}_{i}(x) }  \frac{\delta }{\delta U^{a}_{i}(y) } e^{- \Sigma_\GZ} \right|_{\text{all sources = 0}} \nonumber\\
&=&  \int [\d \Phi] \left[ - g f^{abc} (D_\mu c)^b (x) \varphi_i^c (x) +  ( D \omega_i)^a(x) \right]\nonumber\\
&& \hspace{5cm} ( D \overline \omega_i)^a(y)      e^{- \Sigma_\GZ} \nonumber\\
& =&  \Braket{ - g f^{abc} (D_\mu c)^b (x) \varphi_i^c (x)  ( D \overline \omega_i)^a(y)  }\nonumber\\
&& \hspace{3cm} +   \Braket{( D \omega_i)^a(x)  ( D \overline \omega_i)^a(y)  } \;. \label{aa}
\end{eqnarray}
We shall now show  that the first correlator in expression \eqref{aa} vanishes,
\begin{eqnarray}\label{green}
\Braket{  (g f^{abc} (D_\mu c)^b  \varphi_i^c) (x)  ( D \overline \omega_i)^a(y)  } &=& 0\;.
\end{eqnarray}
In fact, equation \eqref{green} belongs to a more general class of Green functions which are all zero, namely
\begin{eqnarray}\label{green2}
\Braket{  \Theta (x)  \Lambda(y)  } &=& 0\;,
\end{eqnarray}
with $ \Theta (x) $ a function of fields \textit{not} containing the field $\omega$, while $\Lambda(y)$ is a function containing $\overline \omega$. We can show that all these Green functions are zero by using an elementary  diagrammatical argument. It is impossible to construct any diagram which has an $\overline \omega$ leg starting from a space time point $y$ which has to be connected in some way to a space time point $x$, where no $\omega$ leg is present. Indeed, every $\overline \omega$ requires an $\omega$ leg to propagate, $\omega$ in his turn shall always produce another $\overline \omega$ leg in all vertices as can be seen from the action \eqref{SGZecht}.  Moreover,  the field $\overline \omega$ needs again another $\omega$ leg to propagate. Therefore, an $\omega$ leg is required in the space time point $y$ to close the diagram.  As a consequence, all Green functions of the type of equation \eqref{green2} are zero.\\
\\
Hence, we conclude that the Green function \\
$\Braket{( D \omega_i)^a(x)  ( D \overline \omega_i)^a(y)  } $ is multiplicatively renormalizable as follows from
\begin{equation}\label{reshor}
  \Braket{( D \omega_i)^a(x)  ( D \overline \omega_i)^a(y)  }_0 = Z_U^{-1} Z_N^{-1}   \Braket{( D \omega_i)^a(x)  ( D \overline \omega_i)^a(y)  } \;,
\end{equation}
whereby
\begin{eqnarray}
 Z_U^{-1} Z_N^{-1}  &=& Z_A^{1/2}  Z_g=Z_c^{-1} \;,
\end{eqnarray}
see expression \eqref{Z3}.

\subsection{The horizon function $h_1$}
We shall now prove that the first horizon function \eqref{h1} implies a horizon condition which is not multiplicatively renormalizable. In an analogous fashion as in the previous subsection, we can write
\begin{eqnarray}
&&\int \d^d x \Braket{h_1(x)} =  \nonumber\\
&& \gamma^4  \int \d^d x \int \d^d y \Braket{ g^2 f^{akc} A^k_\mu(x) (\mathcal M^{-1})^{ab} (x,y) f^{b\ell c} A^\ell_\mu (y) } \nonumber\\
&&=  - \frac{1}{N_c}\gamma^4  \int \d^d x \int \d^d y  \Braket{ (g f^{akc} A^k_\mu \omega^a_i )(x)  (g  f^{b\ell c} A^\ell_\mu \overline \omega^b_i)  (y) } \;. \nonumber\\
\end{eqnarray}
We can demonstrate that the composite operators \\$g f^{akc} A^k_\mu \omega^a_i $ and $g  f^{b\ell c} A^\ell_\mu \overline \omega^b_i$ are not renormalizable. This is due to the fact that, at the quantum level, those composite operators will unavoidably mix with the operators $\p_\mu \omega^{a}_i$, $\p_\mu \overline\omega^{a}_i$, which have the same quantum numbers. For this, we need to consider the operators $\p_\mu \omega^{a}_i$ and $g f_{akb} A^k_\mu \omega^{b}_i$  as separate operators, each coupled to their own source, instead of $D_\mu^{ab} \omega_i^b$ being coupled to the single source $U_\mu^{a i}$, and similarly for $\p_\mu\overline \omega^{a}_i$ and $g f_{akb} A^k \overline \omega^{b}_i$. In Appendix B, we have given an alternative proof of the renormalizability of the Gribov-Zwanziger action, where we have coupled the following sources to the following operators, see equation \eqref{s2},
\begin{eqnarray}
   \p_\mu \omega_i^a &\rightarrow& U_\mu^{ai} \;, \nonumber\\
     g f^{akb} A_\mu^k \omega^{b}_i-g f_{abc} D_\mu^{bd }c^d \varphi_i^c & \rightarrow&  U_\mu^{\prime ai} \;,\nonumber\\
     \p_\mu \overline \omega_i^a  & \rightarrow& -N_\mu^{ai} \;,  \nonumber\\
      g f^{akb} A_\mu^k \overline \omega^{b}_i  & \rightarrow&   - N^{\prime a i}_{\mu} \;.
\end{eqnarray}
In \eqref{zmatrix1}, we have found that the sources $U$ and $U'$ mix, as well as  $N$ mixes with $N'$. Taking the inverse of this matrix yields
\begin{align}
 \left[
  \begin{array}{c}
    U \\
    U' \\
  \end{array}
\right]&=\left[
          \begin{array}{cc}
               Z_A^{-1/2} & a_1 \\
            0 & Z_g^{1}   \\
          \end{array}
        \right]
\left[
  \begin{array}{c}
    U_0\\
    U'_0
  \end{array}
\right]\,, \nonumber\\
 \left[
  \begin{array}{c}
    N \\
    N' \\
  \end{array}
\right]&=\left[
          \begin{array}{cc}
              Z_g^{-1}  & a_1 \\
            0 &  Z_A^{1/2}  \\
          \end{array}
        \right]
\left[
  \begin{array}{c}
    N_0\\
    N'_0
  \end{array}
\right]\,.
\end{align}
We recall that insertions of an operator can be obtained by taking derivatives of the generating functional \\$Z^c(U,U',N,N')$ w.r.t.~to the appropriate source. For example,
\begin{multline}
(g f^{akb} A_\mu^k \overline \omega^{b}_i)_0 \sim \frac{\delta Z^c((U,U',N,N')}{\delta N'_0}\nonumber\\
     = \frac{\delta N}{\delta N'_0}\frac{\delta Z^c(U,U',N,N') }{\delta N}+\frac{\delta N'}{\delta N'_0}\frac{\delta Z^c(U,U',N,N')}{\delta N }
\end{multline}
so that
\begin{eqnarray}\label{vgl1}
    (g f^{akb} A_\mu^k \overline \omega^{b}_i)_0  &=& a_1 ( \p_\mu \overline \omega^{a}_i ) + Z_A^{1/2} (g f^{akb} A_\mu^k \overline \omega^{b}_i)\;.
\end{eqnarray}
We can do the same for the other operator
\begin{eqnarray}\label{vgl2}
     &&(g f^{akb} A_\mu^k \omega^{b}_i-g f_{abc} D_\mu^{bd }c^d \varphi_i^c)_0  \nonumber\\
     &&= a_1 ( \p_\mu  \omega^{a}_i ) + Z_g (g f^{akb} A_\mu^k \omega^{b}_i - g f_{abc} D_\mu^{bd }c^d \varphi^c_i)\;.
\end{eqnarray}
Taking the partial derivatives of $\Sigma^2_\GZ$ given in equation \eqref{B1} w.r.t.~$ N^{\prime a}_{i}(x)$ and $U^{\prime a}_{i}(y)$, we find
\begin{eqnarray*}\label{2terms}
&&\int [\d \Phi] \left. \frac{\delta }{\delta N^{\prime a}_{i}(x) }  \frac{\delta }{\delta U^{\prime a}_{i}(y) } e^{- \Sigma_\GZ} \right|_{\text{all sources = 0}} \nonumber\\
&&=  \int [\d \Phi] \left(g f^{akb} A_\mu^k \omega^{b}_i-g f_{abc} D_\mu^{bd }c^d \varphi_i^c \right)(x) \nonumber\\
&& \hspace{4cm}\times (g f^{a\ell c} A_\mu^\ell \overline \omega^{ c }_i)(y)      e^{- \Sigma_\GZ} \nonumber\\
&&=  \Braket{ (g f^{akb} A_\mu^k \omega^{b}_i) (x)  (g f^{a\ell c} A_\mu^\ell \overline \omega^{c }_i)(y) } \nonumber\\
&& \hspace{2cm}-   \Braket{ (g f_{abc} D_\mu^{bd }c^d \varphi_i^c) (x)  (g f^{a\ell c} A_\mu^\ell \overline \omega^{c}_i)(y) }\;.
\end{eqnarray*}
The correlator $\Braket{ (g f_{abc} D_\mu^{bd }c^d \varphi_i^c) (x)  (g f^{akb} A_\mu^k \overline \omega^{b}_i)(y) }$ also belongs to the class \eqref{green2} and is therefore equal to zero. However, the remaining Green function  \\$\Braket{ (g f^{akb} A_\mu^k \omega^{b}_i) (x)  (g f^{a\ell c} A_\mu^\ell \overline \omega^{c}_i)(y) }$ is no longer multiplicatively renormalizable. Indeed, due to the mixing we have found, the bare correlator can be written as follows
\begin{eqnarray}\label{renorm}
&&\Braket{ (g f^{akb} A_\mu^k \omega^{b}_i) (x)  (g f^{a\ell c} A_\mu^\ell \overline \omega^{c}_i)(y) }_0 \nonumber\\
&&= a_1 Z_A^{1/2} \Braket{ (\p_\mu \omega^{a}_i) (x)  (g f^{a\ell c} A_\mu^\ell \overline \omega^{ c}_i)(y) } \nonumber\\
&&+ a_1 Z_g \Braket{ (g f^{akb} A_\mu^k \omega^{b}_i) (x)  (\p_\mu \overline \omega^{a}_i)(y) }\nonumber\\
 &&+ Z_g Z_A^{1/2} \Braket{ (g f^{akb} A_\mu^k \omega^{b}_i) (x)  (g f^{a\ell c} A_\mu^\ell \overline \omega^{ c}_i)(y) }\;.
\end{eqnarray}
This is a strong argument why one should not rely on this first horizon $h_1$ function. When renormalizing this Green function, the ``missing" terms stemming from the covariant derivative re-enter again.

\section{The composite operators, part II}
\subsection{The horizon function $h_3$}
Let us now look at the composite operators which are involved in the horizon term $h_3$ after having performed the necessary localization. In particular, we are interested in the composite operators appearing in expression \eqref{hor1a} before setting $\gamma(x)$ equal to a constant, namely $D_\mu^{bc} \varphi^{cb}_{\mu}$ and $D_\mu^{bc} \overline{\varphi}^{bc}_{\mu}$. If the horizon function is well defined, we expect these composite operators to be renormalizable. We can check this again from Appendix A. Firstly, from \eqref{brstinvariant} we can deduce
\begin{eqnarray}
(D_\mu^{ab} \varphi^{b}_i)_0 &=& Z_M^{-1}  (D_\mu^{ab} \varphi^{b}_i ) = Z_g^{1/2} Z_A^{1/4} (D_\mu^{ab} \varphi^{b}_i ) \;.
\end{eqnarray}
Secondly, we can do something analogous for $D_\mu^{ab} \overline \varphi^{b}_i$. We see that this operator is a linear combination of two renormalizable composite operators, namely the composite operator coupled to $V_\mu^{ai}$, i.e.~$-D_\mu^{ab} \overline \varphi^b_i + g f_{abc} D^{bd}_\mu c^d \overline \omega^c_i$ and to $R_\mu^{ai}$, i.e.~$- g f_{abc} D^{bd}_\mu c^d \overline \omega^c_i$. Luckily, the two composite operators have the same renormalization constant, as without this property, the linear combination would not be renormalizable. Therefore,
\begin{eqnarray}
&&(D_\mu^{ab} \overline \varphi^b_i)_0   \nonumber\\
&&=  - (-D_\mu^{ab} \overline \varphi^b_i + g f_{abc} D^{bd}_\mu c^d \overline \omega^c_i)_0 - (- g f_{abc} D^{bd}_\mu c^d \overline \omega^c_i)_0 \nonumber\\
&&= Z_V^{-1} (D_\mu^{ab} \overline \varphi^b_i)\;,
\end{eqnarray}
whereby $Z_V^{-1} = Z_g^{1/2}Z_A^{1/4}$ as can be found in \eqref{Z3}. In conclusion, both composite operators $D_\mu^{bc} \varphi^{cb}_{\mu}$ and $D_\mu^{bc} \overline{\varphi}^{bc}_{\mu}$ are renormalizable.

\subsection{The horizon function $h_1$}
When taking the limit $\theta(x) \to 0$, we can drop the total derivatives. However, the remaining composite operators $g f_{akb} A^k \varphi^{b}_i$ and $g f_{akb} A^k \overline \varphi^{bi}$ are not multiplicative renormalizable. We can prove this in an analogous fashion as in the previous section. In Appendix B, we have considered the operators $\p_\mu \varphi^{a}_i$ and $g f_{akb} A^k_\mu \varphi^{b}_i$  as separate operators instead of being coupled only to one source, and similarly for $\p_\mu\overline \varphi^{a}_i$ and $g f_{akb} A^k \overline \varphi^{b}_i$.  To that purpose, in the Appendix B, we have coupled the following sources to the following operators, see equation \eqref{s2}:
\begin{eqnarray}
   \p_\mu  \varphi^{a}_i &\rightarrow& M_{\mu}^{ai} \;,\nonumber\\
    g f^{akb} A_\mu^k \varphi^{b}_i & \rightarrow& M^{\prime ai}_{\mu} \;,\nonumber\\
     \p_\mu \overline \varphi^{a}_i  & \rightarrow& V_{\mu}^{ai} \;,\nonumber\\
       (g f_{akb} A_\mu^{k} \overline \varphi^{b}_i - g f_{abc} D^{bd}_\mu c^d \overline \omega^c_i)  & \rightarrow&  V_{\mu}^{\prime ai}\;,
\end{eqnarray}
and we have reworked out the complete algebraic renormalization in this alternative setting. In \eqref{zmatrix1}, we have found that the sources $M$ and $M'$ mix, and also $V$ mixes with $V'$. Taking the inverse of this matrix yields
\begin{align}\label{zmatrix1}
 \left[
  \begin{array}{c}
    M \\
    M' \\
  \end{array}
\right]&=\left[
          \begin{array}{cc}
              Z_g^{-1/2} Z_A^{-1/4} & a_1 \\
            0 & Z_g^{1/2} Z_A^{1/4}  \\
          \end{array}
        \right]
\left[
  \begin{array}{c}
    M_0\\
    M'_0
  \end{array}
\right]\,,  \nonumber\\
 \left[
  \begin{array}{c}
    V\\
    V' \\
  \end{array}
\right]&=\left[
         \begin{array}{cc}
              Z_g^{-1/2} Z_A^{-1/4} & a_1 \\
            0 & Z_g^{1/2} Z_A^{1/4}  \\
          \end{array}
        \right]
\left[
  \begin{array}{c}
    V_0\\
    V'_0
  \end{array}
\right]\,.
\end{align}
As in the previous section we have that
\begin{equation}
    (g f^{akb} A_\mu^k \varphi^{b}_i)_0 = a_1 ( \p_\mu  \varphi^{a}_i ) + Z_g^{-1/2} Z_A^{-1/4} (g f^{akb} A_\mu^k \varphi^{b}_i)\;.
\end{equation}
We can do the same for the other operator. Only some care has to be taken as the source $V_\mu^{\prime ai}$ couples to a sum of two operators $ (g f_{akb} A_\mu^{k} \overline \varphi^{b}_i - g f_{abc} D^{bd}_\mu c^d \overline \omega^c_i) $. We therefore have to subtract the operator coupled to the source $R_\mu^{ai}$, namely
\begin{eqnarray}
&&(  g f_{akb} A_\mu^{k} \overline \varphi^{b}_i)_0\nonumber\\
&=&(g f_{akb} A_\mu^{k} \overline \varphi^{b}_i - g f_{abc} D^{bd}_\mu c^d \overline \omega^c_i)_0 - (- g f_{abc} D^{bd}_\mu c^d \overline \omega^c_i)_0 \nonumber\\
  &=& a_1 ( \p_\mu  \overline \varphi_i^{a} ) + Z_g^{-1/2} Z_A^{-1/4} (g f^{akb} A_\mu^k \overline \varphi^{b}_i  - g f_{abc} D^{bd}_\mu c^d \overline \omega^c_i)\nonumber\\
&&   \hspace{2cm}-  Z_g^{1/2} Z_A^{1/4}(- g f_{abc} D^{bd}_\mu c^d \overline \omega^c_i) \nonumber\\
  &=& a_1 ( \p_\mu  \overline \varphi_i^{a} ) + Z_g^{-1/2} Z_A^{-1/4} (g f^{akb} A_\mu^k \overline \varphi^{b}_i  ) \;.
\end{eqnarray}
We can thus conclude that the operators $g f_{akb} A^k \varphi^{b}_i$ and $g f_{akb} A^k \overline \varphi^{b}_i$ are not multiplicatively renormalizable and mix with the operators $\p_\mu \varphi^{a}_i$ and $\p_\mu\overline \varphi^{a}_i$, respectively. Therefore, one should keep in mind that the limit $\theta \to 0$ has to be taken as the final step, and that it can only be taken at the local level. Furthermore, the mixing we have found tells us that one should always leave the covariant derivative of a field ``in one piece''. As a consequence, much care has to be taken at the nonlocal level when deriving all kinds of results, as we shall now explain in the next section.

\section{The Kugo-Ojima Green function(s) and the link with the horizon condition}
As explained in \cite{Dudal:2009xh}, there exists a close link between the Gribov-Zwanziger formalism and the Kugo-Ojima analysis of gauge theories. An important role in the Kugo-Ojima work is played by the parameter $u(0)$, which is the zero momentum limit of the function $u(p^2)$, which can be extracted from
\begin{multline}\label{KO1}
    \int \d^d x e^{ipx}\braket{ D_\mu^{ad}c^d(x) D_\nu^{be}\overline c^e(0)}\\
    =\delta^{ab}\left[\left(\delta_{\mu\nu}-\frac{p_\mu p_\nu}{p^2}\right)u(p^2)-\frac{p_\mu p_\nu}{p^2}\right]\;.
\end{multline}
In order to have a finite quantity $u(p^2)$, the previous Green function should of course be renormalizable. In the case of pure Yang-Mills theories, one can indeed prove this, but this was achieved by also using anti-BRST invariance. We refer to the literature \cite{Dudal:2003dp,Grassi:2004yq} for details. In the presence of the restriction to the Gribov region $\Omega$, we even lack the concept of the anti-BRST symmetry as far as we know. We shall therefore present a slightly different argument here. Using the correspondence \eqref{head}, we may also write
\begin{multline}
    -\frac{1}{N_c}\int \d^d x \e^{ipx}\braket{ D_\mu^{ad}\omega^{d s}_{\lambda}(x) D_\nu^{be}\overline \omega^{e s}_{\lambda }(0)} \\
    =\delta^{ab}\left[\left(\delta_{\mu\nu}-\frac{p_\mu p_\nu}{p^2}\right)u(p^2)-\frac{p_\mu p_\nu}{p^2}\right] \;,
\end{multline}
from which the renormalizability immediately follows, see equation \eqref{reshor}. If one wishes to introduce bare quantities, one can write down
\begin{eqnarray}\label{KO3}
    &&-\frac{1}{N_c}\int \d^d x e^{ipx}\braket{ D_\mu^{ad}\omega^{d s}_{\lambda}(x) D_\nu^{be}\overline \omega^{e s}_{\lambda }(0)}_0\nonumber\\
    &&=\delta^{ab}\left[\left(\delta_{\mu\nu}-\frac{p_\mu p_\nu}{p^2}\right)u(p^2)-\frac{p_\mu p_\nu}{p^2}\right]_0\nonumber\\
    \Leftrightarrow&& -\frac{Z_c^{-1}}{N_c}\int \d^d x e^{ipx}\braket{ D_\mu^{ad}\omega^{d s}_{\lambda}(x) D_\nu^{be}\overline \omega^{e s}_{\lambda }(0)}\nonumber\\
    &&=Z_c^{-1}\delta^{ab}\left[\left(\delta_{\mu\nu}-\frac{p_\mu p_\nu}{p^2}\right)u(p^2)-\frac{p_\mu p_\nu}{p^2}\right]_0\;,
\end{eqnarray}
but this is just a formal relation. It nevertheless teaches us that one cannot simply speak about the (multiplicative) renormalization of $u(p^2)$ in terms of a bare Kugo-Ojima function $u_0(p^2)$.\\\\
As a corollary, using the identification \eqref{ident}, we derive the following correspondence between the horizon function and the Kugo-Ojima parameter $u(0)$,
\begin{equation}\label{KO4}
    -\braket{h_3}= (N^2-1)\left[(d-1)u(0)-1\right]\;,
\end{equation}
which hereby rigourously proves the relation which was one of the starting points of the paper \cite{Dudal:2009xh}.\\\\
In \cite{Kugo:1995km}, it was shown that one can parametrize the ghost propagator, defined as follows
\begin{eqnarray}\label{KO5}
\Braket{c^a(-p) \overline{c}^b(p)}&=\delta^{ab} G (p^2)\;,
\end{eqnarray}
in terms of\footnote{Actually, in \cite{Kugo:1995km}, another notation $v(p^2)$ has been used instead of $w(p^2)$, the relation being  $ w(p^2) = p^2v(p^2)$ .}
\begin{eqnarray}\label{KO6}
G (p^2) &=& \frac{1}{p^2(1 + u(p^2) + w (p^2))}\;.
\end{eqnarray}
This relation was also discussed in \cite{Zwanziger:1992qr,Kondo:2009wk,Kondo:2009gc,Kondo:2009ug,Aguilar:2009nf,Boucaud:2009sd}. Using this relation, one can again derive a formal correspondence between bare and finite quantities,
\begin{eqnarray}\label{KO7}
[1 + u(p^2) + w (p^2)]_0=Z_c^{-1}[1 + u(p^2) + w (p^2)]\;.
\end{eqnarray}
but as already noticed in \cite{Boucaud:2009sd}, it does not allow to separately speak about $(1+u(p^2))_0$ and $w_0(p^2)$.\\\\ In \cite{Kondo:2009wk}, it was claimed that the function $w(p^2)$ enjoyed the property of being not-renormalized. We are unable to understand this claim, especially since we do not know what the corresponding bare quantity $w_0(p^2)$ would be. Let us explain.  Due to expression \eqref{head}, we can rewrite \eqref{ww} as
\begin{eqnarray}\label{KO8}
&&- \int \frac{\d^d x}{N_c} \d^d y e^{i p(x-y) } \Braket{ (g f_{abc} A_\mu^b \omega_i^c) (x)  (g f_{ak\ell} A_\nu^k \overline \omega^\ell_i ) (y) }_{1PI} \nonumber\\
  &&\hspace{3cm}=  g_{\mu\nu} u(p^2) + \frac{p_\mu p_\nu}{p^2} w(p^2)\;.
\end{eqnarray}
We see that this correlator is, up to the Lorentz structure, equal to the one investigated in equation \eqref{renorm}.  Also this correlator is not multiplicatively renormalizable, as we find
\begin{eqnarray}\label{renorm2}
&&\Braket{ (g f^{akb} A_\mu^k \omega^{b}_i) (x)  (g f^{a\ell c} A_\nu^\ell \overline \omega^{c}_i)(y) }_0 \nonumber\\
&=& a_1 Z_A^{1/2} \Braket{ (\p_\mu \omega^{a}_i) (x)  (g f^{a\ell c} A_\nu^\ell \overline \omega^{ c}_i)(y) } \nonumber\\
&&+ a_1 Z_g \Braket{ (g f^{akb} A_\mu^k \omega^{b}_i) (x)  (\p_\nu \overline \omega^{a}_i)(y) }\nonumber\\
 && + Z_g Z_A^{1/2} \Braket{ (g f^{akb} A_\mu^k \omega^{b}_i) (x)  (g f^{a\ell c} A_\nu^\ell \overline \omega^{ c}_i)(y) }\;.
\end{eqnarray}
It is therefore of no use to define $u(p^2)$ and $w(p^2)$ at the bare level, as we cannot introduce their corresponding bare counterparts.  We can conclude that the definition of $u(p^2)$ and $w(p^2)$ should be considered as a definition at the level of renormalized Green functions, in which case the parametrizations \eqref{KO1} and \eqref{KO6} make perfect sense. Due to the mixing, we do not see how we can maintain the definition \eqref{KO8} at the renormalized level. In our opinion, one must first define $u(p^2)$ through equation \eqref{KO1}, which then allows to extract a value for $w(p^2)$ from the ghost propagator via the relation \eqref{KO6}.\\
\\
We end by commenting on the results of \cite{Kondo:2009wk,Kondo:2009gc,Kondo:2009ug} which gave a connection between the horizon function $h_1$ given in equation \eqref{h1} and the functions $u(p^2)$ and $w(p^2)$. Using the nonlocal expression \eqref{h1}, one manages to write down a quite complicated connection. However, as we have clearly shown in this work, it is impossible to treat $h_1$ at the quantum level, as quantum effects make necessary the presence of the derivative operators, see the previous section. At the end, only the horizon function $h_3$ given in equation \eqref{danhorizon} is meaningful, and as such only the connection \eqref{KO4} survives. In contrast with the outcome of\footnote{The author of \cite{Kondo:2009wk} came to the conclusion that a straightforward implementation of the horizon condition leads to $G(0)=3$, rather than $G(0)=\infty$.} \cite{Kondo:2009wk}, which turns out to be ill-defined, imposing the horizon condition \eqref{horizoncondition} would lead to the infrared enhanced ghost as follows from the results \eqref{KO4}, \eqref{KO6}, and using the fact that  $w(0)=0$. This has been confirmed explicitly in \cite{Gracey:2005cx} to two loop order, and it has also been found using a Schwinger-Dyson approach in \cite{Aguilar:2009nf}.\\\\The question that remains then is whether the ghost propagator is necessarily infrared enhanced when enforcing the horizon condition? Indeed, starting from the standard Gribov-Zwanziger action, one shall find an enhanced ghost propagator. As already indicated in the introduction, this does not seem to be in agreement with recent lattice results. However, after one takes into account the nontrivial dynamics of the Gribov-Zwanziger action by including the possibility of having nonvanishing condensates, one is able to find different results \cite{Dudal:2007cw,Dudal:2008sp,Dudal:2010tf}. By including for example the dimension 2 condensate $\braket{\overline \varphi^{a}_i \varphi^{a}_i -\overline \omega^{a}_i \omega^{a}_i}$, a non-enhanced ghost propagator is found within the so called refined Gribov-Zwanziger (RGZ) formalism  \cite{Dudal:2007cw,Dudal:2008sp,Dudal:2010tf}. Simultaneously, the same condensate also leads to a gluon propagator that does not vanish at zero momentum. Moreover, we mention that the form of the gluon propagator stemming from the  refined Gribov-Zwanziger framework enables us to fit the lattice data in a very reasonable way \cite{Dudal:2010tf}. We notice that the works \cite{Kondo:2009wk,Kondo:2009gc,Kondo:2009ug} are unable to say anything about the gluon sector.

\section{Conclusion}
The main conclusion of this paper is that the renormalizability requirement of the Green functions of a quantum field theory is of paramount importance. It dictates whether the Green function one is considering makes sense or not. Once having obtained a finite, renormalized Green function, one can address the question of how the renormalized Green function can be parametrized in terms of certain quantities. It is of no use, or even incorrect, to discuss the renormalization of such parameters directly.\\\\
An important ingredient to discuss and interpret the underlying renormalizable quantum field theory is the issue of locality. If one wishes to study nonlocal theories, a prerequisite appears to be that one can write down a local version of the theory, to which the usual tools of local quantum field theory apply \cite{Capri:2007ck}. Once having established the necessary features of a quantum field theory at the local level, one can try to go back and see what the corresponding nonlocal version looks like.\\\\ In particular in this paper we have first shown that the Gribov-Zwanziger action is unambiguous. There is only one possible horizon function at the nonlocal level, which is given by
\begin{multline}
\int d^d x h(x)  \equiv \lim_{\theta \to 0}  \int d^d x h_3(x) = \lim_{\theta \to 0}  \int \d^d y \\
\; \left( D_\mu^{ac}(x) \gamma^2(x) \right) (\mathcal M^{-1})^{ab}(x,y) ( D_\mu^{bc} (y) \gamma^2(y) ) \;.
\end{multline}
with $\gamma^2 (z) = \e^{i \theta z} \gamma^2$. We have motivated in two different ways that this is the only correct horizon function to start any discussion from. Our analysis was based on the
renormalization properties of the corresponding local version of the operators involved.\\\\
In addition, we have also elaborated on the definitions of the Kugo-Ojima parameters $u(p^2)$ and $w(p^2)$, and we have argued  that it only makes sense to talk about these parameters at the level of renormalized Green functions. We reanalyzed the connection between $u(p^2)$ and the horizon function, and motivated that the relation \eqref{KO4} is the only correct one, in contrast with other results made available recently \cite{Kondo:2009wk}.  We believe that any relation derived from other horizon functions should be looked upon with great caution \cite{Kondo:2009wk}.

\section*{Acknowledgments}
We have benefitted from useful discussions with K.~I.~Kondo, J.~Rodriguez-Quintero and D.~Zwanziger. We would also like to thank A. Defauw and Markus Q. Huber for a careful reading of the manuscript. D.~Dudal and N.~Vandersickel are supported by the Research Foundation-Flanders (FWO). The Conselho Nacional de Desenvolvimento Cien\-t\'{\i}fico e Tecnol\'{o}gico (CNPq-Brazil), the Faperj,  Funda{\c{c}}{\~{a}}o de Amparo {\`{a}} Pesquisa do Estado do Rio de Janeiro, the SR2-UERJ and the Coordena{\c{c}}{\~{a}}o de
 Aperfei{\c{c}}oamento de Pessoal de N{\'{\i}}vel Superior (CAPES) are gratefully acknowledged for financial support. D.~Dudal and N.~Vandersickel wish to thank the UERJ for its hospitality during the final stages of this work.

\appendix
\section{Renormalization of the Gribov-Zwanziger action, option 1}
In this section we shall repeat the proof of the renormalizability of the Gribov-Zwanziger action. We shall also have the opportunity to correct a minor  mistake present in  previous works \cite{Dudal:2005na,Dudal:2008sp,Dudal:2009zh}. This is the reason why we shall repeat the full algebraic renormalization in all details. In particular, our current analysis also presents a slightly improved version of the original renormalization proof \cite{Zwanziger:1992qr,Maggiore:1993wq}. However, let us stress that none of our results are affected by this minor error, like e.g.~those presented in \cite{Dudal:2008sp,Dudal:2009zh}.

\subsection{The starting action and the BRST}
We start with the following action,
\begin{eqnarray}\label{start}
S_\GZ &=& S_{0} +  S_{\gamma} \,,
\end{eqnarray}
with
\begin{eqnarray}
S_{0}&=&S_\YM + S_\gf + \int \d^d x \left( \overline \varphi_i^a \p_\mu \left( D_\mu^{ab} \varphi^b_i \right)  \right.\nonumber\\
&&\left. \hspace{1cm}- \overline \omega_i^a \p_\mu \left( D_\mu^{ab} \omega_i^b \right) - g f^{abc} \p_\mu \overline \omega_i^a    D_\mu^{bd} c^d  \varphi_i^c \right) \nonumber \;, \\
S_{\gamma}&=& -\gamma ^{2}g\int\d^{d}x\left( f^{abc}A_{\mu }^{a}\varphi _{\mu }^{bc} \right. \nonumber\\
&&\left. \hspace{2cm} +f^{abc}A_{\mu}^{a}\overline{\varphi }_{\mu }^{bc} + \frac{d}{g}\left(N^{2}-1\right) \gamma^{2} \right) \;.
\end{eqnarray}
First, notice that we have simplified the notation of the additional fields $\left( \overline \varphi_\mu^{ac},\varphi_\mu^{ac},\overline \omega_\mu^{ac},\omega_\mu^{ac}\right) $ in $S_0$ as $S_0$ displays a symmetry with respect to the composite index $i=\left( \mu,c\right)$. Therefore, we have set
\begin{equation}
\left( \overline \varphi_\mu^{ac},\varphi_\mu^{ac},\overline \omega_\mu^{ac},\omega_\mu^{ac}\right) =\left( \overline \varphi_i^a,\varphi_i^a,\overline \omega_i^a,\omega_i^a \right)\,.
\end{equation}
The conventional BRST variations of all the fields are given by,
\begin{align}\label{BRST1}
sA_{\mu }^{a} &=-\left( D_{\mu }c\right) ^{a}\,, & sc^{a} &=\frac{1}{2}gf^{abc}c^{b}c^{c}\,,   \nonumber \\
s\overline{c}^{a} &=b^{a}\,,&   sb^{a}&=0\,,  \nonumber \\
s\varphi _{i}^{a} &=\omega _{i}^{a}\,,&s\omega _{i}^{a}&=0\,,\nonumber \\
s\overline{\omega}_{i}^{a} &=\overline{\varphi }_{i}^{a}\,,& s \overline{\varphi }_{i}^{a}&=0\,.
\end{align}
One can check that the BRST symmetry is softly broken for the Gribov-Zwanziger action \cite{Zwanziger:1989mf,Dudal:2008sp},
\begin{eqnarray*}
&&s S_\GZ = s (S_{0} +  S_{\gamma}) ~=~ s (  S_{\gamma}) \nonumber\\
&&~=~ g \gamma^2 \int \d^d x f^{abc} \left( A^a_{\mu} \omega^{bc}_\mu -
 \left(D_{\mu}^{am} c^m\right)\left( \overline{\varphi}^{bc}_\mu + \varphi^{bc}_{\mu}\right)  \right)\,.
\end{eqnarray*}
We notice that the breaking is due to the $\gamma$ dependent term, $S_\gamma$.\\
\\
In order to discuss the renormalizability of $S_\GZ$, we should treat the breaking as a composite operator to be introduced into the action by means of a suitable set of external sources. This procedure can be done in a BRST invariant way, by embedding $S_\GZ$ into a larger action, namely
\begin{eqnarray}\label{brstinvariant}
\Sigma_\GZ &=& S_{\YM} + S_{\gf} + S_0 + S_\s \,,
\end{eqnarray}
whereby
\begin{eqnarray}\label{previous}
S_\s &=& s\int \d^d x \left( -U_\mu^{ai} D_\mu^{ab} \varphi_i^b - V_\mu^{ai} D_{\mu}^{ab} \overline \omega_i^{b} - U_\mu^{ai} V_\mu^{ai} \right.\nonumber\\
&& \left.\hspace{5cm} + T_\mu^{a i} g f_{abc} D^{bd}_\mu c^d \overline \omega^c_i \right)\nonumber\\
&=& \int \d^d x \left( -M_\mu^{ai}  D_\mu^{ab} \varphi_i^b - gf^{abc} U_\mu^{ai}   D^{bd}_\mu c^d  \varphi_i^c \right. \nonumber\\
&&  + U_\mu^{ai}  D_\mu^{ab} \omega_i^b - N_\mu^{ai}  D_\mu^{ab} \overline \omega_i^b - V_\mu^{ai}  D_\mu^{ab} \overline \varphi_i^b \nonumber\\
&&+ gf^{abc} V_\mu^{ai} D_\mu^{bd} c^d \overline \omega_i^c - M_\mu^{ai} V_\mu^{ai}+U_\mu^{ai} N_\mu^{ai} \nonumber \\
&& \left. + R_\mu^{ai} g f^{abc} D_\mu^{bd} c^d \overline \omega^c_i  + T_\mu^{ai} g f_{abc} D^{bd}_\mu c^d \overline \varphi^c_i\right) \,.
\end{eqnarray}
We have introduced 3 new doublets ($U_\mu^{ai}$, $M_\mu^{ai}$), ($V_\mu^{ai}$, $N_\mu^{ai}$) and ($T_\mu^{ai}$, $R_\mu^{ai}$) with the following BRST transformations,
and
\begin{align}\label{BRST2}
sU_{\mu }^{ai} &= M_{\mu }^{ai}\,, & sM_{\mu }^{ai}&=0\,,  \nonumber \\
sV_{\mu }^{ai} &= N_{\mu }^{ai}\,, & sN_{\mu }^{ai}&=0\,,\nonumber \\
sT_{\mu }^{ai} &= R_{\mu }^{ai}\,, & sR_{\mu }^{ai}&=0\;.
\end{align}
We have therefore restored the broken BRST at the expense of introducing new sources. However, we do not want to alter our original theory \eqref{SGZ}. Therefore, at the end, we have to set the sources equal to the following values:
\begin{eqnarray}\label{physlimit}
&& \left. U_\mu^{ai}\right|_{\phys} = \left. N_\mu^{ai}\right|_{\phys} = \left. T_\mu^{ai}\right|_{\phys} = 0 \,, \nonumber\\
&& \left. M_{\mu \nu }^{ab}\right|_{\phys}= \left.V_{\mu \nu}^{ab}\right|_{\phys}=  -\left.R_{\mu \nu}^{ab}\right|_{\phys} = \gamma^2 \delta ^{ab}\delta _{\mu \nu } \,.
\end{eqnarray}
It is exactly here that we have committed a mistake. In previous papers, we have forgotten to introduce the doublet ($T_\mu^{ai}$, $R_\mu^{ai}$), and therefore, the intended physical limit did not exactly reproduce the original action \eqref{start}. In the original article \cite{Zwanziger:1992qr}, these two sources were also not introduced. When taking the physical limit, an extra term was generated, which was then removed by doing a (nonlocal) shift in the $\omega$ field. Here, we have circumvented this unnecessary shift by introducing the doublet ($T_\mu^{ai}$, $R_\mu^{ai}$).

\subsection{The Ward identities}
Following the procedure of algebraic renormalization \cite{Piguet:1995er}, we should try to find all possible Ward identities. Before doing this, in order to be able to write the Slavnov-Taylor identity, we first have to couple all nonlinear BRST transformations to a new source. Looking at \eqref{BRST1}, we see that only $A_\mu^a$ and $c^a$ transform nonlinearly under the BRST $s$. Therefore, we add the following term to the action $\Sigma_\GZ $,
\begin{eqnarray}\label{ext}
S_{\mathrm{ext}}&=&\int \d^4x\left( -K_{\mu }^{a}\left( D_{\mu }c\right) ^{a}+\frac{1}{2}gL^{a}f^{abc}c^{b}c^{c}\right) \;,
\end{eqnarray}
with $K_{\mu }^{a}$ and $L^a$ two new sources which shall be put to zero at the end,
\begin{eqnarray}
\left. K_{\mu }^{a}\right|_{\phys} =\left. L^{a}\right|_{\phys}  = 0\;.&
\end{eqnarray}
These sources are invariant under the BRST transformation,
\begin{align}
s K_{\mu }^{a} &=0\;, & s L^{a} &= 0\;.
\end{align}
The new action is therefore given by
\begin{eqnarray}
\Sigma'_\GZ &=& \Sigma_\GZ + S_{\mathrm{ext}} \;.
\end{eqnarray}
The next step is now to find all the Ward identities obeyed by the action $\Sigma'_\GZ$. We have enlisted all the identities below:

\begin{table}
\caption{Quantum numbers of the fields.}
\label{tabel1}
\begin{tabular}{|c|c|c|c|c|c|c|c|c|}
\hline
& $A_{\mu }^{a}$ & $c^{a}$ & $\overline{c}^{a}$ & $b^{a}$ & $\varphi_{i}^{a} $ & $\overline{\varphi }_{i}^{a}$ &                $\omega _{i}^{a}$ & $\overline{\omega }_{i}^{a}$   \\
\hline
\textrm{dimension} & $1$ & $0$ &$2$ & $2$ & $1$ & $1$ & $1$ & $1$ \\
$\mathrm{ghost\; number}$ & $0$ & $1$ & $-1$ & $0$ & $0$ & $0$ & $1$ & $-1$ \\
$Q_{f}\textrm{-charge}$ & $0$ & $0$ & $0$ & $0$ & $1$ & $-1$& $1$ & $-1$ \\
\hline
\end{tabular}
\end{table}

\begin{table}
\begin{center}
\caption{Quantum numbers of the sources.}
\label{tabel2}
\begin{tabular}{|c|c|c|c|c|c|c|c|c|}\hline
        $U_{\mu}^{ai}$&$M_{\mu }^{ai}$&$N_{\mu }^{ai}$&$V_{\mu }^{ai}$& $R_{\mu }^{ai}$  &  $T_{\mu }^{ai}$ &$K_{\mu }^{a}$&$L^{a}$  \\
\hline
         $2$ & $2$ & $2$ &$2$ & 2&2  & $3$ & $4$  \\
         $-1$& $0$ & $1$ & $0$ & 0& -1 & $-1$ & $-2$  \\
         $-1$ & $-1$ & $1$ & $1$ &1&1& $0$ & $0$  \\
\hline
\end{tabular}
\end{center}
\end{table}

\begin{enumerate}
\item The Slavnov-Taylor identity is given by
\begin{equation}\label{slavnov}
\mathcal{S}(\Sigma'_\GZ )=0\;,
\end{equation}
with
\begin{multline*}
\mathcal{S}(\Sigma'_\GZ ) =\int \d^4x\left( \frac{\delta \Sigma'_\GZ
}{\delta K_{\mu }^{a}}\frac{\delta \Sigma'_\GZ }{\delta A_{\mu
}^{a}}+\frac{\delta \Sigma'_\GZ }{\delta L^{a}}\frac{\delta \Sigma'_\GZ
}{\delta c^{a}} \right. \nonumber\\
+b^{a}\frac{\delta \Sigma'_\GZ
}{\delta \overline{c}^{a}}+\overline{\varphi }_{i}^{a}\frac{\delta \Sigma'_\GZ }{\delta \overline{\omega }_{i}^{a}}+\omega _{i}^{a}\frac{\delta \Sigma'_\GZ }{\delta \varphi _{i}^{a}} \nonumber\\
\left. +M_{\mu }^{ai}\frac{\delta \Sigma'_\GZ}{\delta U_{\mu}^{ai}}+N_{\mu }^{ai}\frac{\delta \Sigma'_\GZ }{\delta V_{\mu }^{ai}} + R_{\mu }^{ai}\frac{\delta \Sigma'_\GZ }{\delta T_{\mu }^{ai}}\right) \;.
\end{multline*}

\item The $U(f)$ invariance is given by
\begin{equation}
U_{ij} \Sigma'_\GZ =0\;,\label{ward1}
\end{equation}
\begin{multline}
U_{ij}=\int \d^dx\Bigl( \varphi_{i}^{a}\frac{\delta }{\delta \varphi _{j}^{a}}-\overline{\varphi}_{j}^{a}\frac{\delta }{\delta \overline{\varphi}_{i}^{a}}+\omega _{i}^{a}\frac{\delta }{\delta \omega _{j}^{a}}-\overline{\omega }_{j}^{a}\frac{\delta }{\delta \overline{\omega }_{i}^{a}} \\
-  M^{aj}_{\mu} \frac{\delta}{\delta M^{ai}_{\mu}} -U^{aj}_{\mu}\frac{\delta}{\delta U^{ai}_{\mu}} + N^{ai}_{\mu}\frac{\delta}{\delta N^{aj}_{\mu}}    \\
  +V^{ai}_{\mu}\frac{\delta}{\delta V^{aj}_{\mu}}    +  R^{aj}_{\mu}\frac{\delta}{\delta R^{ai}_{\mu}} + T^{aj}_{\mu}\frac{\delta}{\delta T^{ai}_{\mu}} \Bigr)  \;. \nonumber
\end{multline}
By means of the diagonal operator $Q_{f}=U_{ii}$, the
$i$-valued fields and sources turn out to possess an additional quantum number.
One can find all quantum numbers in Table \ref{tabel1} and Table \ref{tabel2}.

\item The Landau gauge condition reads
\begin{eqnarray}
\frac{\delta \Sigma'_\GZ }{\delta b^{a}}&=&\partial_\mu A_\mu^{a}\;.
\end{eqnarray}

\item The antighost equation yields
\begin{eqnarray}
\frac{\delta \Sigma'_\GZ }{\delta \overline{c}^{a}}+\partial _{\mu}\frac{\delta \Sigma'_\GZ }{\delta K_{\mu }^{a}}&=&0\;.
\end{eqnarray}

\item The linearly broken local constraints yield
\begin{eqnarray}
\frac{\delta \Sigma'_\GZ }{\delta \overline{\varphi }^{a}_i}+\partial _{\mu }\frac{\delta \Sigma'_\GZ }{\delta M_{\mu }^{ai}} + g f_{dba}    T^{d i}_\mu \frac{\delta \Sigma'_\GZ }{\delta K_{\mu }^{b i}} =gf^{abc}A_{\mu }^{b}V_{\mu}^{ci} \;, \nonumber\\
\frac{\delta \Sigma'_\GZ }{\delta \omega ^{a}_i}+\partial _{\mu}\frac{\delta \Sigma'_\GZ }{\delta N_{\mu}^{ai}}-gf^{abc}\overline{\omega }^{b}_i\frac{\delta \Sigma'_\GZ }{\delta b^{c}}=gf^{abc}A_{\mu }^{b}U_{\mu }^{ci}\nonumber\\ \;.
\end{eqnarray}

\item The exact $\mathcal{R}_{ij}$ symmetry reads
\begin{equation}
\mathcal{R}_{ij}\Sigma'_\GZ =0\;,
\end{equation}
with
\begin{multline}\label{rij}
\mathcal{R}_{ij} = \int \d^4x\left( \varphi _{i}^{a}\frac{\delta}{\delta\omega _{j}^{a}}-\overline{\omega }_{j}^{a}\frac{\delta }{\delta \overline{\varphi }_{i}^{a}}\right.\\
\left.+V_{\mu }^{ai}\frac{\delta }{\delta N_{\mu}^{aj}}-U_{\mu }^{aj}\frac{\delta }{\delta M_{\mu }^{ai}} + T^{a i}_\mu \frac{\delta }{\delta R_{\mu }^{aj}}  \right) \;.
\end{multline}

\item The integrated Ward identity is given by
\begin{equation}
\int \d^4 x \left( c^a \frac{ \delta \Sigma'_\GZ }{ \delta \omega^{ a}_i} + \overline \omega^{a}_i \frac{ \delta \Sigma'_\GZ }{ \delta  \overline c^a } + U^{a i}_\mu \frac{ \delta \Sigma'_\GZ }{ \delta  K^a_\mu }  \right) = 0\;.
\end{equation}
\end{enumerate}
Here we should add that due to the presence of the sources $T_{\mu }^{ai}$ and $R_{\mu }^{ai}$, the ghost Ward identity \cite{Piguet:1995er} is broken, and we are unable to restore this identity. For the standard Yang-Mills theory, this identity has the following form
\begin{equation}\label{GW}
\mathcal{G}^{a}( S_\YM + S_\gf)=\Delta _{\mathrm{cl}}^{a}\,,
\end{equation}
with
\begin{eqnarray}
\mathcal{G}^{a} &=&\int \d^dx\left( \frac{\delta }{\delta c^{a}}+gf^{abc} \overline{c}^{b}\frac{\delta }{\delta b^{c}} \right) \,,
\end{eqnarray}
and
\begin{equation}
\Delta _{\mathrm{cl}}^{a}=g\int \d^{4}xf^{abc}\left( K_{\mu}^{b}A_{\mu }^{c}-L^{b}c^{c}\right) \;,
\end{equation}
a linear breaking. However, it shall turn out that this is not a problem for the renormalization procedure being undertaken.

\subsection{The counterterm}
The next step in the algebraic renormalization is to translate all these symmetries into constraints on the counterterm $\Sigma_\GZ^c$, which is an integrated polynomial in the fields and sources of dimension four and with ghost number zero. The classical action $\Sigma_\GZ'$ changes under quantum corrections according to
\begin{eqnarray}\label{deformed}
    \Sigma_\GZ' \rightarrow \Sigma_\GZ' + h \Sigma_\GZ^c\,,
\end{eqnarray}
whereby $h$ is the perturbation parameter. Demanding that the perturbed action $ (\Sigma_\GZ' + h \Sigma_\GZ^c) $ fulfills the same set of Ward identities obeyed by $\Sigma_\GZ'$, see \cite{Piguet:1995er}, it follows that the counterterm $\Sigma_\GZ^c$ is constrained by the following identities.
\begin{enumerate}
\item The linearized Slavnov-Taylor identity yields
\begin{equation}
\mathcal{B}\Sigma_\GZ^{c}=0\;,
\end{equation}
with $\mathcal{B}$ the nilpotent linearized Slavnov-Taylor operator,
\begin{multline}
\mathcal{B}=\int \d^{4}x\Bigl( \frac{\delta \Sigma_\GZ'}{\delta K_{\mu }^{a}}\frac{\delta }{\delta A_{\mu }^{a}}+\frac{\delta \Sigma_\GZ' }{\delta A_{\mu }^{a}}\frac{\delta }{\delta K_{\mu }^{a}}+\frac{\delta \Sigma_\GZ' }{\delta L^{a}}\frac{\delta }{\delta c^{a}}\\
+\frac{\delta\Sigma_\GZ' }{\delta c^{a}}\frac{\delta }{\delta L^{a}}+b^{a}\frac{\delta }{\delta \overline{c}^{a}}+\overline{\varphi}_{i}^{a}\frac{\delta }{\delta \overline{\omega }_{i}^{a}}+\omega_{i}^{a}\frac{\delta }{\delta \varphi_{i}^{a}}\\
+M_{\mu }^{ai}\frac{\delta }{\delta U_{\mu }^{ai}} + N_{\mu }^{ai}\frac{\delta }{\delta V_{\mu }^{ai}} + R_{\mu }^{ai}\frac{\delta }{\delta T_{\mu }^{ai}}  \Bigr) \,,
\end{multline}
and
\begin{equation}
\mathcal{B}^2=0\;.
\end{equation}

\item The $U(f)$ invariance reads
\begin{eqnarray}
U_{ij} \Sigma_\GZ^{c} &=&0 \;.
\end{eqnarray}

\item The Landau gauge condition
\begin{eqnarray}
\frac{\delta \Sigma_\GZ^{c}}{\delta b^{a}}&=&0\,.
\end{eqnarray}

\item The antighost equation
\begin{eqnarray}
\frac{\delta \Sigma_\GZ^{c}}{\delta \overline c^{a}}+\p_\mu\frac{\delta \Sigma_\GZ^{c}}{\delta K_{\mu}^a} &=&0 \,.
\end{eqnarray}

\item The linearly broken local constraints yield
\begin{multline}
\left( \frac{\delta  }{\delta \overline{\varphi }^{a}_i}+\partial _{\mu }\frac{\delta}{\delta M_{\mu }^{ai}} +\partial _{\mu }\frac{\delta }{\delta M_{\mu }^{ai}} + g f_{abc}    T^{b i}_\mu \frac{\delta }{\delta K_{\mu }^{c i}}\right) \Sigma_\GZ^{c} \\=0 \;,
\end{multline}
\begin{multline}
\left( \frac{\delta }{\delta \omega ^{a}_i}+\partial _{\mu}\frac{\delta }{\delta N_{\mu}^{ai}}-gf^{abc}\overline{\omega }^{b}_i \frac{\delta }{\delta b^{c}} \right) \Sigma_\GZ^{c}\\
=0 \;.
\end{multline}

\item The exact $\mathcal{R}_{ij}$ symmetry reads
\begin{equation}
\mathcal{R}_{ij}\Sigma_\GZ^{c}=0\;,
\end{equation}
with  $\mathcal{R}_{ij}$ given in \eqref{rij}.

\item Finally, the integrated Ward identity becomes
\begin{equation}
\int \d^4 x \left( c^a \frac{ \delta \Sigma_\GZ^{c}}{ \delta \omega^{ a}_i} + \overline \omega^{a}_i \frac{ \delta \Sigma_\GZ^{c}}{ \delta  \overline c^a } + U^{a i}_\mu \frac{ \delta \Sigma_\GZ^{c}}{ \delta  K^a_\mu }  \right) = 0 \;.
\end{equation}
\end{enumerate}
Now we can write down the most general counterterm $\Sigma_\GZ^{c}$ of $d=4$, which obeys the linearized Slavnov-Taylor identity, has ghost number zero, and vanishing $Q_f$ number,
\begin{multline}\label{counterterm}
\Sigma^c_\GZ = a_0 S_{\YM} + \mathcal{B} \int \d^d \!x\,   \biggl\{ \biggl[ a_{1} K_{\mu}^{a} A_{\mu}^{a} + a_2 \partial _{\mu} \overline{c}^{a} A_{\mu}^{a}+a_3 \,L^{a}c^{a} \\
+a_4 U_{\mu}^{ai}\,\partial _{\mu }\varphi _{i}^{a} +a_5 \,V_{\mu}^{ai}\,\partial _{\mu }\overline{\omega }_{i}^{a} +a_6\,\overline{\omega }_{i}^{a}\partial ^{2}\varphi _{i}^{a}+a_7\, \,U_{\mu}^{ai}V_{\mu}^{ai}\\
+a_8\,gf^{abc}U_{\mu}^{ai}\,\varphi _{i}^{b}A_{\mu }^{c}+a_9\,gf^{abc}V_{\mu}^{ai}\,\overline{\omega }_{i}^{b}A_{\mu }^{c}\\
+a_{10}\,gf^{abc}\overline{\omega }_{i}^{a}A_{\mu }^{c}\,\partial _{\mu }\varphi _{i}^{b} +a_{11}\,gf^{abc}\overline{\omega }_{i}^{a}(\partial _{\mu }A_{\mu}^{c})\varphi _{i}^{b}  \\
+ b_1 R_{\mu}^{ai} U_{\mu}^{ai} +b_2 T_{\mu }^{ai} M_{\mu }^{ai} + b_3 g f_{abc} R_{\mu }^{ai} \overline{\omega }_{i}^{b} A_{\mu}^{c}\\
 + b_4 g f_{abc} T_{\mu}^{ai} \overline{\varphi }_{i}^{b} A_{\mu}^{c} + b_5 R_{\mu}^{ai} \p_\mu \overline{\omega }_{i}^{a}  + b_6 T_{\mu}^{ai} \p_\mu \overline{\varphi }_{i}^{a} \biggr] \biggr\} \;,
\end{multline}
with $a_0, \ldots, a_{11}$ arbitrary parameters. Now we can unleash the constraints on the counterterm. Firstly, although the the ghost Ward identity \eqref{GW} is broken, we know that this is not so in the standard Yang-Mills case. Therefore, we can already set $a_3=0$ as this term is not allowed in the counterterm of the standard Yang-Mills action, which is a special case of the action we are studying\footnote{In particular, since we will always assume the use of a mass independent renormalization scheme, we may compute $a_3$ with all external mass scales (= sources) equal to zero. Said otherwise, $a_3$ is completely determined by the dynamics of the original Yang-Mills action, in which case it is known to vanish to all orders \cite{Piguet:1995er}.}. Secondly, due to the Landau gauge condition (3.) and the antighost equation (4.) we find,
\begin{eqnarray}
a_1 &=& a_2\;.
\end{eqnarray}
Next, the linearly broken constraints (5.) give the following relations
\begin{align}
 a_1 &= -a_8  = - a_{9} = a_{10} = a_{11} = -b_3 = b_4\;, \nonumber\\
  a_4 &= a_5 = -a_6 = a_7\;, \quad b_1 =b_2 = b_5 = b_6 = 0 \;.
\end{align}
The $R_{ij}$ symmetry does not give any new information, while the integrated Ward identity relates the two previous strings of parameters:
\begin{multline}
 a_1 = -a_8  = - a_{9} = a_{10} = a_{11} = -b_3 = b_4  \\ \equiv     a_3 = a_4 = -a_5 = a_6 \;.
\end{multline}
Taking all this information together, we obtain the following counterterm
\begin{multline}\label{final}
\Sigma^c= a_{0}S_{YM}  +
a_{1}\int \d^dx\Biggl(  A_{\mu}^{a}\frac{ \delta S_{YM}}{\delta A_{\mu }^{a}}  + \p_\mu \overline{c}^a \p_\mu c^a  \\
+ K_{\mu }^{a}\partial _{\mu }c^{a}  + M_\mu^{a i} \p_\mu \varphi^{a}_i -  U_\mu^{a i} \p_\mu \omega^{a}_i + N_\mu^{a i} \p_\mu \overline{\omega}_i^{a} +  V_\mu^{a i}\p_\mu \overline{\varphi}^{a}_i\\
 +  \p_\mu \overline{\varphi}^{a}_i \p_\mu \varphi^{a}_i +  \p_\mu \omega^{a}_i \p_\mu \overline{\omega}^{a}_i + V_\mu^{a i} M_\mu^{a i} - U_\mu^{a i}N_\mu^{a i}  \\
- g f_{abc} U_\mu^{ai} \varphi^{b}_i \p_\mu c^c - g f_{abc} V_\mu^{ai} \overline{\omega}^{b}_i \p_\mu c^c - g f_{abc} \p_{\mu} \overline{\omega}^a_i \varphi^{b}_i  \p_\mu c^c \nonumber\\
  - g f_{abc} R_\mu^{ai} \p_\mu c^b \overline \omega_i^c + g f_{abc} T_\mu^{ai} \p_\mu c^b \overline \varphi_i^c \Biggr) \;.
\end{multline}

\subsection{The renormalization factors}
As a final step, we have to show that the counterterm \eqref{final} can be reabsorbed by means of a multiplicative renormalization of the fields and sources.
If we try to absorb the counterterm into the original action, we easily find,
\begin{eqnarray}\label{Z1}
Z_{g} &=&1-h \frac{a_0}{2}\,,  \nonumber \\
Z_{A}^{1/2} &=&1+h \left( \frac{a_0}{2}+a_{1}\right) \,,
\end{eqnarray}
and
\begin{eqnarray}\label{Z2}
Z_{\overline{c}}^{1/2} &=& Z_{c}^{1/2} = Z_A^{-1/4} Z_g^{-1/2} = 1-h \frac{a_{1}}{2}\,, \nonumber \\
Z_{b}&=&Z_{A}^{-1}\,, \nonumber\\
Z_{K }&=&Z_{c}^{1/2}\,,  \nonumber\\
Z_{L} &=&Z_{A}^{1/2}\,.
\end{eqnarray}
The results \eqref{Z1} are already known from the renormalization of the original Yang-Mills action in the Landau gauge. Further, we also obtain
\begin{eqnarray}\label{Z3}
Z_{\varphi}^{1/2} &=& Z_{\overline \varphi}^{1/2} = Z_g^{-1/2} Z_A^{-1/4} = 1 - h \frac{a_1}{2}\,, \nonumber\\
Z_\omega^{1/2} &=& Z_A^{-1/2} \,,\nonumber\\
Z_{\overline \omega}^{1/2} &=& Z_g^{-1} \,,\nonumber\\
Z_M &=& 1- \frac{a_1}{2} = Z_g^{-1/2} Z_A^{-1/4}\,, \nonumber\\
Z_N &=& Z_A^{-1/2} \,, \nonumber\\
Z_U &=& 1 + h \frac{a_0}{2} = Z_g^{-1} \,, \nonumber\\
Z_V &=& 1- h \frac{a_1}{2} = Z_g^{-1/2}Z_A^{-1/4} \,,  \nonumber\\
Z_T &=& 1+h \frac{a_0}{2} = Z_g^{-1}  \,,  \nonumber\\
Z_R &=& 1- h \frac{a_1}{2} = Z_g^{-1/2}Z_A^{-1/4}\;.
\end{eqnarray}
This concludes the proof of the renormalizability of the action \eqref{start} which is the physical limit of $\Sigma_\GZ' $.

\section{Renormalization of the Gribov-Zwanziger action, option 2}
We can also write down an alternative proof of the renormalization of the Gribov-Zwanziger action. Looking at \eqref{h1bis}, we are tempted to
treat $g f_{akb} A^k_\mu \varphi^{bc}_\nu$ and $g f_{akb} A^k_\mu \overline \varphi^{bc}_\nu$ as the relevant composite operators. However, we shall see that these operators mix with $\p_\mu \varphi^{ac}_\nu$ and $\p_\mu\overline \varphi^{ac}_\nu$, respectively. A similar observation holds for $g f_{akb} A_\mu^k \omega^{bc}_\nu$ and $g f_{akb} A_\mu^k \overline \omega^{bc}_\nu$. Such conclusion is evidently not unexpected, as we have learnt already from Appendix A that e.g.~\\
$D_\mu^{ab}\varphi_\nu^{bc}=\p_\mu \varphi^{ab}_\nu+g f_{akb} A^k_\mu \varphi^{bc}_\nu$ is multiplicatively renormalizable.

\subsection{The starting action and the BRST}
Looking at  \eqref{brstinvariant}, we may try to also start with another possible BRST invariant action,
\begin{eqnarray}\label{B1}
\Sigma_\GZ^{(2)} &=& S_{\YM} + S_{\gf} + S_0 + S_\s^{(2)} + S_{\mathrm{ext}} \,,
\end{eqnarray}
whereby we have immediately added $S_{\mathrm{ext}}$ still given by \eqref{ext} and with
\begin{eqnarray}\label{s2}
S_\s^{(2)} &=& s\int \d^d x \left( -U_\mu^{ai} \p_\mu \varphi_i^a - U_\mu^{\prime ai} g f_{akb} A^k_\mu \varphi_i^b - V_\mu^{ai} \p_{\mu} \overline \omega_i^{a} \right. \nonumber\\
 && \left. - V_\mu^{\prime ai} g f_{akb} A^k_\mu \overline \omega_i^b      - U_\mu^{\prime ai} V_\mu^{ \prime ai}  + T_\mu^{a i} g f_{abc} D^{bd}_\mu c^d \overline \omega^c_i \right)\nonumber\\
&=& \int \d^d x \left( -M_\mu^{ai}  \p_\mu  \varphi_i^a   + U_\mu^{ai}  \p_\mu \omega_i^a  - M^{\prime a i}_{\mu} g f^{akb} A_\mu^k \varphi^{b}_i   \right. \nonumber\\
&& - gf^{abc} U_\mu^{\prime ai}   D^{bd}_\mu c^d  \varphi_i^c + U_\mu^{\prime ai}  g f_{akb} A_\mu^{k} \omega_i^b -N_\mu^{ai}  \p_\mu \overline \omega_i^a  \nonumber\\
  && + V_\mu^{ai}  \p_\mu \overline \varphi_i^a  - N^{\prime a i}_{\mu} g f^{akb} A_\mu^k \overline \omega^{b i}  - gf^{abc} V_\mu^{\prime ai}   D^{bd}_\mu c^d  \overline \omega_i^c  \nonumber \\
&& + V_\mu^{\prime ai}  g f_{akb} A_\mu^{k} \overline \varphi_i^b + R_\mu^{ai} g f^{abc} D_\mu^{bd} c^d \overline \omega^c_i \nonumber\\
 && \left. + T_\mu^{ai} g f_{abc} D^{bd}_\mu c^d \overline \varphi^c_i\right) \,.
\end{eqnarray}
In contrast with the previous section, we have now introduced 5 doublets, ($U_\mu^{ai}$, $M_\mu^{ai}$), ($U_\mu^{\prime ai}$, $M_\mu^{\prime ai}$), ($V_\mu^{ai}$, $N_\mu^{ai}$), ($V_\mu^{\prime ai}$, $N_\mu^{\prime ai}$) and ($T_\mu^{ai}$, $R_\mu^{ai}$) with the following BRST transformations,
\begin{align}
sU_{\mu }^{ai} &= M_{\mu }^{ai}\,, & sM_{\mu }^{ai}&=0\,,  \nonumber \\
sU_{\mu }^{\prime ai} &= M_{\mu }^{\prime ai}\,, & sM_{  \mu }^{\prime ai}&=0\,,  \nonumber \\
sV_{\mu }^{ai} &= N_{\mu }^{ai}\,, & sN_{\mu }^{ai}&=0\,,\nonumber \\
sV_{\mu }^{\prime ai} &= N_{\mu }^{\prime ai}\,, & sN_{\mu }^{\prime ai}&=0\,,\nonumber \\
sT_{\mu }^{ai} &= R_{\mu }^{ai}\,, & sR_{\mu }^{ai}&=0\;.
\end{align}
In order to go back from $S_\s^{(2)}$ to $S_\s$ from Appendix A (see equation \eqref{previous}), we just need to set $U = U'$, $V= V'$, $N = N'$ and $M= M'$. Eventually, it appears natural to give
the primed sources the same physical value of their corresponding unprimed counterparts, see equation \eqref{physlimit}.

\subsection{The Ward identities}
Just as in Appendix A, we enlist all the Ward identities obeyed by $\Sigma_\GZ^{(2)}$, which of course look very similar.

\begin{enumerate}
\item The Slavnov-Taylor identity is now given by
\begin{equation}
\mathcal{S}(\Sigma^{(2)}_\GZ )=0\;,
\end{equation}
with
\begin{eqnarray*}
&&\mathcal{S}(\Sigma^{(2)}_\GZ ) =\int \d^4x\Bigl( \frac{\delta \Sigma^{(2)}_\GZ }{\delta K_{\mu }^{a}}\frac{\delta \Sigma^{(2)}_\GZ }{\delta A_{\mu }^{a}}+\frac{\delta \Sigma^{(2)}_\GZ }{\delta L^{a}}\frac{\delta \Sigma^{(2)}_\GZ }{\delta c^{a}} \nonumber\\
&& +b^{a}\frac{\delta \Sigma^{(2)}_\GZ }{\delta \overline{c}^{a}}+\overline{\varphi }_{i}^{a}\frac{\delta \Sigma^{(2)}_\GZ }{\delta \overline{\omega }_{i}^{a}}+\omega _{i}^{a}\frac{\delta \Sigma^{(2)}_\GZ }{\delta \varphi _{i}^{a}} + R_{\mu }^{ai}\frac{\delta \Sigma^{(2)}_\GZ }{\delta T_{\mu }^{ai}}\nonumber\\
&&+M_{\mu }^{ai}\frac{\delta \Sigma^{(2)}_\GZ }{\delta U_{\mu}^{ai}}+N_{\mu }^{ai}\frac{\delta \Sigma^{(2)}_\GZ }{\delta V_{\mu }^{ai}} +M_{\mu }^{\prime ai}\frac{\delta \Sigma^{(2)}_\GZ
}{\delta U_{\mu}^{\prime ai}}+N_{\mu }^{\prime ai}\frac{\delta \Sigma^{(2)}_\GZ }{\delta V_{\mu }^{\prime ai}} \Bigr) \;.
\end{eqnarray*}

\item The $U(f)$ invariance is easily adapted
\begin{equation}
U_{ij} \Sigma^{(2)}_\GZ =0\;,
\end{equation}
\begin{multline}
U_{ij}=\int \d^dx\Bigl( \varphi_{i}^{a}\frac{\delta }{\delta \varphi _{j}^{a}}-\overline{\varphi}_{j}^{a}\frac{\delta }{\delta \overline{\varphi}_{i}^{a}}+\omega _{i}^{a}\frac{\delta }{\delta \omega _{j}^{a}} -\overline{\omega }_{j}^{a}\frac{\delta }{\delta \overline{\omega }_{i}^{a}}\\
-  M^{aj}_{\mu} \frac{\delta}{\delta M^{ai}_{\mu}} -  M^{\prime aj}_{\mu} \frac{\delta}{\delta M^{\prime ai}_{\mu}} -U^{aj}_{\mu}\frac{\delta}{\delta U^{ai}_{\mu}} - U^{\prime aj}_{\mu}\frac{\delta}{\delta U^{\prime ai}_{\mu}}    \nonumber\\
+ N^{ai}_{\mu}\frac{\delta}{\delta N^{aj}_{\mu}}+ N^{\prime ai}_{\mu}\frac{\delta}{\delta N^{\prime aj}_{\mu}} +V^{ai}_{\mu}\frac{\delta}{\delta V^{aj}_{\mu}}  +V^{\prime ai}_{\mu}\frac{\delta}{\delta V^{\prime aj}_{\mu}} \\
 +  R^{aj}_{\mu}\frac{\delta}{\delta R^{ai}_{\mu}} + T^{aj}_{\mu}\frac{\delta}{\delta T^{ai}_{\mu}} \Bigr)  \;. \nonumber
\end{multline}
We have again that the $i$-valued fields and sources turn out to possess an additional quantum number. All the quantum number are still the same as in Table \ref{tabel1} and Table \ref{tabel2}, whereby we keep in mind that the quantum numbers of the primed sources are obviously the same as those of the unprimed ones.

\item The Landau gauge condition does not change,
\begin{eqnarray}
\frac{\delta \Sigma^{(2)}_\GZ }{\delta b^{a}}&=&\partial_\mu A_\mu^{a}\;.
\end{eqnarray}

\item The same goes for the antighost equation,
\begin{eqnarray}
\frac{\delta \Sigma^{(2)}_\GZ }{\delta \overline{c}^{a}}+\partial _{\mu}\frac{\delta \Sigma^{(2)}_\GZ }{\delta K_{\mu }^{a}}&=&0\;.
\end{eqnarray}

\item The linearly broken local constraints now become
\begin{multline}
\frac{\delta \Sigma^{(2)}_\GZ }{\delta \overline{\varphi }^{a}_i}+\partial _{\mu }\frac{\delta \Sigma^{(2)}_\GZ }{\delta M_{\mu }^{ai}}  +\partial _{\mu }\frac{\delta \Sigma^{(2)}_\GZ }{\delta M_{\mu }^{ \prime ai}} \\+ g f_{dba}    T^{d i}_\mu \frac{\delta \Sigma^{(2)}_\GZ }{\delta K_{\mu }^{b i}} =gf^{abc}A_{\mu }^{b}V_{\mu}^{\prime ci} \;,
\end{multline}
\begin{multline}
\frac{\delta \Sigma^{(2)}_\GZ }{\delta \omega^{a}_i}+\partial _{\mu}\frac{\delta \Sigma^{(2)}_\GZ }{\delta N_{\mu}^{ai}} +\partial _{\mu}\frac{\delta \Sigma^{(2)}_\GZ }{\delta N_{\mu
}^{\prime ai}} \\-gf^{abc}\overline{\omega }^{bi}\frac{\delta \Sigma^{(2)}_\GZ }{\delta b^{c}}=gf^{abc}A_{\mu }^{b}U_{\mu }^{\prime ci} \;.
\end{multline}
We also find some extra linearly broken constraints
\begin{align}
\frac{\delta \Sigma^{(2)}_\GZ }{\delta M_{\mu}^{ai}} &= \p_\mu \varphi^{ai}    \;,&
\frac{\delta \Sigma^{(2)}_\GZ }{\delta N_{\mu}^{ai}} &= \p_\mu \overline \omega^{ai}  \;, \nonumber\\
\frac{\delta \Sigma^{(2)}_\GZ }{\delta U_{\mu}^{ai}} &= \p_\mu \omega^{ai}   \;,&
\frac{\delta \Sigma^{(2)}_\GZ }{\delta V_{\mu}^{ai}} &= \p_\mu  \overline \varphi^{ai}   \;.
\end{align}

\item The exact $\mathcal{R}_{ij}$ symmetry can be adapted to
\begin{equation}
\mathcal{R}_{ij}\Sigma^{(2)}_\GZ =0\;,
\end{equation}
with
\begin{eqnarray}
&&\mathcal{R}_{ij} = \int \d^4x\Bigl( \varphi _{i}^{a}\frac{\delta}{\delta\omega _{j}^{a}}-\overline{\omega }_{j}^{a}\frac{\delta }{\delta \overline{\varphi }_{i}^{a}}+V_{\mu }^{ai}\frac{\delta }{\delta N_{\mu}^{ai}} \nonumber\\
&& +V_{\mu }^{\prime ai}\frac{\delta }{\delta N_{\mu}^{\prime aj}}-U_{\mu }^{aj}\frac{\delta }{\delta M_{\mu }^{ai}} -U_{\mu }^{\prime ai}\frac{\delta }{\delta M_{\mu }^{\prime ai}} + T^{a i}_\mu \frac{\delta }{\delta R_{\mu }^{aj}}  \Bigr) \;. \nonumber\\
\end{eqnarray}

\item The integrated Ward identity is now linearly broken as follows
\begin{eqnarray}
&&\int \d^4 x \left( c^a \frac{ \delta \Sigma^{(2)}_\GZ }{ \delta \omega^{ a }_i} + \overline \omega^{a}_i \frac{ \delta \Sigma^{(2)}_\GZ }{ \delta  \overline c^a } + U^{\prime a i}_\mu \frac{ \delta \Sigma^{(2)}_\GZ }{ \delta  K^a_\mu }  \right)\nonumber\\
&=& U^{ai}_\mu \p_\mu c^a - U^{\prime ai}_\mu \p_\mu c^a \;.
\end{eqnarray}
\end{enumerate}

\subsection{The counterterm}
We again translate all the identities into identities for the counterterm $\Sigma^{(2)c}_\GZ$
\begin{enumerate}
\item The linearized Slavnov-Taylor identity:
\begin{equation}
\mathcal{B}^{(2)}\Sigma_\GZ^{(2)c}=0\;,
\end{equation}
with $\mathcal{B}^{(2)}$ the nilpotent linearized Slavnov-Taylor operator,
\begin{eqnarray}
&&\mathcal{B}^{(2)} =\int \d^{4}x\Bigl( \frac{\delta \Sigma^{(2)}_\GZ}{\delta K_{\mu }^{a}}\frac{\delta }{\delta A_{\mu }^{a}}+\frac{\delta \Sigma^{(2)}_\GZ }{\delta A_{\mu }^{a}}\frac{\delta }{\delta K_{\mu }^{a}}+\frac{\delta \Sigma^{(2)}_\GZ }{\delta L^{a}}\frac{\delta }{\delta c^{a}}\nonumber\\
&&+\frac{\delta\Sigma^{(2)}_\GZ }{\delta c^{a}}\frac{\delta }{\delta L^{a}}+b^{a}\frac{\delta }{\delta \overline{c}^{a}}+\overline{\varphi}_{i}^{a}\frac{\delta }{\delta \overline{\omega }_{i}^{a}}+\omega_{i}^{a}\frac{\delta }{\delta \varphi_{i}^{a}}+M_{\mu }^{ai}\frac{\delta }{\delta U_{\mu }^{ai}}\nonumber\\
&& + N_{\mu }^{ai}\frac{\delta }{\delta V_{\mu }^{ai}}  +M_{\mu }^{\prime ai}\frac{\delta }{\delta U_{\mu }^{\prime  ai}} + N_{\mu }^{\prime ai}\frac{\delta }{\delta V_{\mu }^{\prime ai}}   + R_{\mu }^{ai}\frac{\delta }{\delta T_{\mu }^{ai}}  \Bigr) \nonumber\,.
\end{eqnarray}

\item The $U(f)$ invariance
\begin{eqnarray}
U_{ij} \Sigma^{(2)c}_\GZ &=&0 \;.
\end{eqnarray}

\item The Landau gauge condition
\begin{eqnarray}
\frac{\delta \Sigma^{(2)c}_\GZ}{\delta b^{a}}&=&0\,.
\end{eqnarray}

\item The antighost equation
\begin{eqnarray}
\frac{\delta \Sigma^{(2)c}_\GZ}{\delta \overline c^{a}}+\p_\mu\frac{\delta \Sigma^{(2)c}_\GZ}{\delta K_{\mu}^a} &=&0 \,.
\end{eqnarray}

\item The linearly broken local constraints
\begin{eqnarray}
&& \left( \frac{\delta  }{\delta \overline{\varphi }^{a}_i}+\partial _{\mu }\frac{\delta}{\delta M_{\mu }^{ai}} +  \partial _{\mu }\frac{\delta  }{\delta M_{\mu }^{ \prime ai}}+ g f_{abc}    T^{b i}_\mu \frac{\delta }{\delta K_{\mu }^{c i}}\right) \nonumber\\
&&\hspace{4cm}\times\Sigma^{(2)c}_\GZ  =0 \;, \nonumber\\
&&\left( \frac{\delta }{\delta \omega^{a}_i}+\partial _{\mu}\frac{\delta }{\delta N_{\mu}^{ai}} +  +\partial _{\mu }\frac{\delta  }{\delta N_{\mu }^{ \prime ai}} -gf^{abc}\overline{\omega }^{b}_i\frac{\delta }{\delta b^{c}} \right) \nonumber\\
&& \hspace{4cm}\times \Sigma^{(2)c}_\GZ  =0 \;,
\end{eqnarray}
and
\begin{align}\label{bigc}
\frac{\delta \Sigma^{(2)c}_\GZ}{\delta M_{\mu}^{ai}} &= 0 \;,&
\frac{\delta \Sigma^{(2)c}_\GZ}{\delta N_{\mu}^{ai}} &= 0\;,\nonumber\\
\frac{\delta \Sigma^{(2)c}_\GZ}{\delta U_{\mu}^{ai}} &= 0\;,&
\frac{\delta \Sigma^{(2)c}_\GZ}{\delta V_{\mu}^{ai}} &= 0\;.
\end{align}
\item The exact $\mathcal{R}_{ij}$ symmetry
\begin{equation}
\mathcal{R}_{ij}\Sigma^{(2)c}_\GZ=0\;.
\end{equation}

\item Finally, the integrated Ward identity becomes
\begin{equation}
\int \d^4 x \left( c^a \frac{ \delta \Sigma^{(2)c}_\GZ}{ \delta \omega^{ a}_i} + \overline \omega^{a}_i \frac{ \delta \Sigma^{(2)c}_\GZ}{ \delta  \overline c^a } + U^{\prime a i}_\mu \frac{ \delta \Sigma^{(2)c}_\GZ}{ \delta  K^a_\mu }  \right) = 0 \;.
\end{equation}
\end{enumerate}
Now we can write down the most general counterterm $\Sigma^{(2)c}_\GZ$ of $d=4$, which obeys the linearized Slavnov-Taylor identity, has ghost number zero, and vanishing $Q_f$ number,
\begin{multline}
\Sigma^{(2)c}_\GZ= a_0 S_{\YM} + \mathcal{B}^{(2)} \int \d^d x \biggl\{ \biggl[ a_{1} K_{\mu}^{a} A_{\mu}^{a} + a_2 \partial _{\mu} \overline{c}^{a} A_{\mu}^{a}+a_3 L^{a}c^{a}\\
 +a_4 U_{\mu}^{ai}\,\partial _{\mu }\varphi_i^a +a_5 V_{\mu}^{ai}\,\partial _{\mu }\overline{\omega }_{i}^{a} +a_6\,\overline{\omega }_{i}^{a}\partial ^{2}\varphi _{i}^{a} +a_7 U_{\mu}^{ai}V_{\mu }^{ai}\nonumber \\
+a_8 gf^{abc}U_{\mu }^{ai}\,\varphi _{i}^{b} A_{\mu }^{c}+a_9 gf^{abc}V_{\mu }^{ai}\,\overline{\omega }_{i}^{b}A_{\mu }^{c}\\
+a_{10} gf^{abc}\overline{\omega }_{i}^{a}A_{\mu }^{c}\,\partial _{\mu }\varphi _{i}^{b} +a_{11}\,gf^{abc}\overline{\omega }_{i}^{a}(\partial _{\mu }A_{\mu}^{c})\varphi _{i}^{b} \\
+ b_1 R_{\mu}^{ai} U_{\mu}^{ai}+b_2 T_{\mu}^{ai} M_{\mu }^{ai}   + b_3 g f_{abc} R_{\mu}^{ai} \overline{\omega }_{i}^{b} A_{\mu}^{c} \\
+ b_4 g f_{abc} T_{\mu}^{ai} \overline{\varphi }_{i}^{b} A_{\mu}^{c} + b_5 R_{\mu}^{ai} \p \overline{\omega }_{i}^{a} + b_6 T_{\mu}^{ai} \p \overline{\varphi }_{i}^{a} +a_4^\prime U_{\mu }^{\prime  ai}\,\partial _{\mu }\varphi _{i}^{a} \\
+a_5^\prime  \,V_{\mu}^{\prime ai}\,\partial _{\mu }\overline{\omega }_{i}^{a}  +a_6^\prime \overline{\omega }_{i}^{\prime a}\partial ^{2}\varphi _{i}^{a} \nonumber\\
+a_7^\prime  \,U_{\mu}^{\prime  ai}V_{\mu}^{\prime ai}+a_8^\prime \,gf^{abc}U_{\mu}^{\prime ai}\,\varphi _{i}^{b}A_{\mu }^{c}+a_9^\prime \,gf^{abc}V_{\mu}^{\prime ai}\,\overline{\omega }_{i}^{b}A_{\mu }^{c}\\
+a_{10}^\prime \,gf^{abc}\overline{\omega }_{i}^{a}A_{\mu }^{c}\,\partial _{\mu }\varphi _{i}^{b} +a_{11}^\prime \,gf^{abc}\overline{\omega }_{i}^{a}(\partial _{\mu }A_{\mu}^{c})\varphi _{i}^{b}  \biggr] \biggr\} \;.
\end{multline}
Notice that the part in $a$ and $b$ parameters is exactly the same as in the previous Appendix A, see equation \eqref{counterterm}. We shall now impose all the constraints induced by the Ward identities. We keep in mind that the argument concerning the broken ghost Ward identity still holds. Also, the 4 constraints \eqref{bigc} invoke the counterterm to be independent of the sources $U'$, $V'$, $M'$ and $N'$. Ultimately, we find
\begin{eqnarray}
&&\Sigma^{(2)c}_\GZ= a_{0}S_{YM}  +
a_{1}\int \d^dx\Biggl(  A_{\mu}^{a}\frac{ \delta S_{YM}}{\delta A_{\mu }^{a}}  + \p_\mu \overline{c}^a \p_\mu c^a \nonumber\\
&&+ K_{\mu }^{a}\partial _{\mu }c^{a}  + M_\mu^{\prime a i} \p_\mu \varphi^{a}_i -  U_\mu^{\prime a i} \p_\mu \omega^{a}_i + N_\mu^{\prime a i} \p_\mu \overline{\omega}^{a}_i \nonumber\\
&&+  V_\mu^{\prime a i} \p_\mu \overline{\varphi}_\mu^{ai}  +  \p_\mu \overline{\varphi}^{a}_i \p_\mu \varphi_i^{a} +  \p_\mu \omega^{a}_i \p_\mu \overline{\omega}^{a}_i + V_\mu^{\prime a i} M_\mu^{\prime a i} \nonumber\\
&&- U_\mu^{\prime a i}N_\mu^{\prime a i} - g f_{abc} U_\mu^{\prime ia} \varphi^{b}_i \p_\mu c^c - g f_{abc} V_\mu^{\prime ia} \overline{\omega}^{b}_i \p_\mu c^c \nonumber\\
&&- g f_{abc} \p_{\mu} \overline{\omega}^a \varphi^{b}_i  \p_\mu c^c  - g f_{abc} R^{ai}_\mu \p_\mu c^b \overline \omega^c + g f_{abc} T^{ai}_\mu \p_\mu c^b \overline \varphi^c \Biggr) \;.\nonumber\\
\end{eqnarray}
We notice the close similarity between this counterterm and the one in expression \eqref{final}.

\subsection{The renormalization factors}
The last step is to find all the renormalization factors. Due to the close similarity with the output of Appendix A, many $Z$ factors will be the same. One can indeed check that equations \eqref{Z1} and \eqref{Z2} still hold, and also the $Z$-factors of $Z_{\varphi}^{1/2}$, $Z_{\overline \varphi}^{1/2} $, $Z_\omega^{1/2}$, $Z_{\overline \omega}^{1/2} $, $Z_T$ and $Z_R$  do not change. Only the renormalization of the sources $U$, $V$, $M$, $N$ is different as they mix with respectively $U'$, $V'$, $M'$, $N'$. Indeed, we find that
\begin{align}
 \left[
  \begin{array}{c}
    M_0 \\
    M'_0 \\
  \end{array}
\right]&=\left[
          \begin{array}{cc}
              Z_g^{1/2} Z_A^{1/4} & -a_1 \\
            0 & Z_g^{-1/2} Z_A^{-1/4}  \\
          \end{array}
        \right]
\left[
  \begin{array}{c}
    M\\
    M'
  \end{array}
\right]\,,  \nonumber\\
 \left[
  \begin{array}{c}
    U_0 \\
    U'_0 \\
  \end{array}
\right]&=\left[
          \begin{array}{cc}
               Z_A^{1/2} & -a_1 \\
            0 & Z_g^{-1}   \\
          \end{array}
        \right]
\left[
  \begin{array}{c}
    U\\
    U'
  \end{array}
\right]\,,  \nonumber\\
 \left[
  \begin{array}{c}
    N_0 \\
    N'_0 \\
  \end{array}
\right]&=\left[
          \begin{array}{cc}
              Z_g^{1}  & -a_1 \\
            0 &  Z_A^{-1/2}  \\
          \end{array}
        \right]
\left[
  \begin{array}{c}
    N\\
    N'
  \end{array}
\right]\,,  \nonumber\\
 \left[
  \begin{array}{c}
    V_0 \\
    V'_0 \\
  \end{array}
\right]&=\left[
          \begin{array}{cc}
               Z_g^{1/2}Z_A^{1/4} & -a_1 \\
            0 & Z_g^{-1/2}Z_A^{-1/4}   \\
          \end{array}
        \right]
\left[
  \begin{array}{c}
    V\\
    V'
  \end{array}
\right]\,,
\end{align}
which again proves the renormalizability of the Gribov-Zwanziger action. The consequences of this mixing shall be explained in sections II.B and III.B.


\begin{thebibliography}{10}
\bibitem{Gribov:1977wm}
  V.~N.~Gribov,
  Nucl.\ Phys.\ B {\bf 139} (1978) 1.

\bibitem{Zwanziger:1989mf}
D.~Zwanziger, Nucl.\ Phys.\  B {\bf 323} (1989) 513.

\bibitem{Dell'Antonio:1989jn}
G.~Dell'Antonio and D.~Zwanziger, Nucl.\ Phys.\  B {\bf 326} (1989) 333.

\bibitem{Zwanziger:1992qr}
D.~Zwanziger, Nucl.\ Phys.\  B {\bf 399} (1993) 477.

\bibitem{Zwanziger:2001kw}
D.~Zwanziger, Phys.\ Rev.\  D {\bf 65} (2002) 094039.

\bibitem{Gracey:2005cx}
J.~A.~Gracey, Phys.\ Lett.\  B {\bf 632} (2006) 282

\bibitem{Ford:2009ar}
F.~R.~Ford and J.~A.~Gracey, J.\ Phys.\ A  {\bf 42} (2009) 325402.

\bibitem{Kondo:2009wk}
K.~I.~Kondo, arXiv:0909.4866 [hep-th].

\bibitem{Kondo:2009gc}
K.~I.~Kondo, arXiv:0907.3249 [hep-th].

\bibitem{Kondo:2009ug}
K.~I.~Kondo, Phys.\ Lett.\  B {\bf 678} (2009) 322.


\bibitem{Cucchieri:2007md}
A.~Cucchieri and T.~Mendes, PoS {\bf LAT2007} (2007) 297

\bibitem{Cucchieri:2008fc}
A.~Cucchieri and T.~Mendes, Phys.\ Rev.\  D {\bf 78} (2008) 094503.

\bibitem{Bornyakov:2008yx}
V.~G.~Bornyakov, V.~K.~Mitrjushkin and M.~Muller-Preussker, Phys.\ Rev.\  D {\bf 79} (2009) 074504.

\bibitem{Bogolubsky:2009dc}
I.~L.~Bogolubsky, E.~M.~Ilgenfritz, M.~Muller-Preussker and A.~Sternbeck, Phys.\ Lett.\  B {\bf 676} (2009) 69.

\bibitem{Maas:2008ri}
A.~Maas, Phys.\ Rev.\  D {\bf 79} (2009) 014505.

\bibitem{Maas:2009se}
A.~Maas, arXiv:0907.5185 [hep-lat].

\bibitem{Maas:2009ph}
A.~Maas, J.~M.~Pawlowski, D.~Spielmann, A.~Sternbeck and L.~von Smekal, arXiv:0912.4203 [hep-lat].

\bibitem{Fischer:2008uz}
C.~S.~Fischer, A.~Maas and J.~M.~Pawlowski, Annals Phys.\  {\bf 324} (2009) 2408.

\bibitem{Oliveira:2007tr}
O.~Oliveira and P.~J.~Silva, PoS {\bf LAT2007} (2007) 332.

\bibitem{Silva:2006av}
P.~J.~Silva and O.~Oliveira, AIP Conf.\ Proc.\  {\bf 892} (2007) 220.


\bibitem{Dudal:2007cw}
D.~Dudal, S.~P.~Sorella, N.~Vandersickel and H.~Verschelde, Phys.\ Rev.\  D {\bf 77} (2008) 071501.

\bibitem{Dudal:2008sp}
D.~Dudal, J.~A.~Gracey, S.~P.~Sorella, N.~Vandersickel and H.~Verschelde, Phys.\ Rev.\  D {\bf 78} (2008) 065047.

\bibitem{Boucaud:2008ji}
Ph.~Boucaud, J.~P.~Leroy, A.~L.~Yaouanc, J.~Micheli, O.~Pene and J.~Rodriguez-Quintero, JHEP {\bf 0806} (2008) 012.

\bibitem{Boucaud:2008ky}
Ph.~Boucaud, J.~P.~Leroy, A.~Le Yaouanc, J.~Micheli, O.~Pene and J.~Rodriguez-Quintero, JHEP {\bf 0806} (2008) 099.

\bibitem{Aguilar:2008xm}
A.~C.~Aguilar, D.~Binosi and J.~Papavassiliou, Phys.\ Rev.\  D {\bf 78} (2008) 025010.

\bibitem{Aguilar:2009nf}
A.~C.~Aguilar, D.~Binosi, J.~Papavassiliou and J.~Rodriguez-Quintero, Phys.\ Rev.\  D {\bf 80} (2009) 085018.

\bibitem{Huber:2009tx}
M.~Q.~Huber, R.~Alkofer and S.~P.~Sorella, Phys.\ Rev.\  D {\bf 81} (2010) 065003

\bibitem{Kugo:1979gm}
T.~Kugo and I.~Ojima, Prog.\ Theor.\ Phys.\ Suppl.\  {\bf 66} (1979) 1.

\bibitem{Kugo:1995km}
T.~Kugo, arXiv:hep-th/9511033.

\bibitem{Dudal:2005na}
D.~Dudal, R.~F.~Sobreiro, S.~P.~Sorella and H.~Verschelde, Phys.\ Rev.\  D {\bf 72} (2005) 014016.

\bibitem{Dudal:2009zh}
D.~Dudal, S.~P.~Sorella, N.~Vandersickel and H.~Verschelde, JHEP {\bf 0908} (2009) 110.

\bibitem{Zwanziger:1993dh}
D.~Zwanziger, Nucl.\ Phys.\  B {\bf 412} (1994) 657.

\bibitem{Dudal:2009xh}
D.~Dudal, S.~P.~Sorella, N.~Vandersickel and H.~Verschelde, Phys.\ Rev.\  D {\bf 79} (2009) 121701.

\bibitem{Dudal:2003dp}
D.~Dudal , V.~E.~R.~Lemes, M.~S.~Sarandy, S.~P.~Sorella, M.~Picariello, A.~Vicini, J.~A.~Gracey and H.~Verschelde,
JHEP {\bf 0306} (2003) 003.

\bibitem{Grassi:2004yq}
P.~A.~Grassi, T.~Hurth and A.~Quadri, Phys.\ Rev.\  D {\bf 70} (2004) 105014.

\bibitem{Boucaud:2009sd}
Ph.~Boucaud, J.~P.~Leroy, A.~Le Yaouanc, J.~Micheli, O.~Pene and J.~Rodriguez-Quintero, Phys.\ Rev.\  D {\bf 80} (2009) 094501.

\bibitem{future}
D.~Dudal, S.~P.~Sorella, N.~Vandersickel, H.~Verschelde, \emph{work in progress}.

\bibitem{Dudal:2010tf}
 D.~Dudal, O.~Oliveira and N.~Vandersickel, arXiv:1002.2374 [hep-lat].

\bibitem{Capri:2007ck}
M.~A.~L.~Capri, V.~E.~R.~Lemes, R.~F.~Sobreiro, S.~P.~Sorella and R.~Thibes, Annals Phys.\  {\bf 323} (2008) 752.

\bibitem{Maggiore:1993wq}
N.~Maggiore and M.~Schaden, Phys.\ Rev.\  D {\bf 50} (1994) 6616.

\bibitem{Piguet:1995er}
O.~Piguet and S.~P.~Sorella, Lect.\ Notes Phys.\ \textbf{M28} (1995) 1.





\end{thebibliography}
\end{document}